\documentclass[10pt,journal,compsoc,twoside]{IEEEtran}

\usepackage[T1]{fontenc}
\usepackage{graphicx}
\usepackage{wrapfig}
\usepackage{float}
\usepackage{graphicx,fancyvrb}
\usepackage{epstopdf}
\usepackage{amsmath,amssymb,amsfonts}
\usepackage{bm}
\usepackage{algorithm}
\newcommand{\algparbox}[1]{\parbox[t]{\dimexpr\linewidth-\algorithmicindent}{#1\strut}}
\usepackage{algpseudocode}
\algnewcommand\algorithmicinput{\textbf{Input:}}
\algnewcommand\algorithmicoutput{\textbf{Output:}}
\algnewcommand\Input{\item[\algorithmicinput]}%
\algnewcommand\Output{\item[\algorithmicoutput]}%
\usepackage{bm}
\usepackage{textcomp}
\usepackage{array,booktabs,makecell}
\def\arraystretch{1.5}
\usepackage{subfigure}
\usepackage{enumerate}
\usepackage{enumitem}
\usepackage{float}
\DeclareMathAlphabet\mathbfcal{OMS}{cmsy}{b}{n}
\usepackage{hyperref}
\AppendGraphicsExtensions{.tif}

\usepackage{tcolorbox}
\usepackage{varwidth}
\usepackage{efbox}
\usepackage{adjustbox}
\usepackage{framed}

\usepackage{etoolbox}

\makeatletter

\long\def\@IEEEtitleabstractindextextbox#1{\parbox{0.922\textwidth}{#1}}

\pretocmd\@bibitem{\csname keycolor#1\endcsname}{}{\fail}
\newcommand\citecolor[3][1]{\@namedef{keycolor#3}{\hspace*{-\labelwidth}\hspace*{-\labelsep}{\color{#2}\rule[-0.3em]{\dimexpr\linewidth+\labelwidth+\labelsep\relax}{#1\baselineskip}}\vspace*{\itemsep}\vspace*{-#1\baselineskip}}}

\makeatother

\ifCLASSOPTIONcompsoc
  \usepackage[nocompress]{cite}
\else
  \usepackage{cite}
\fi

\usepackage[capitalise,noabbrev]{cleveref}
\usepackage{color}
\usepackage[normalem]{ulem}

\usepackage{acronym}
\acrodef{ML}{Machine Learning}
\acrodef{AI}{Artificial Intelligence}
\acrodef{RSS}{Received Signal Strength}
\acrodef{KPI}{Key Performance Indicator}
\acrodef{QoS}{Quality of Service}
\acrodef{SPM}{Standard Propagation Model}
\acrodef{SUI}{Stanford University Interim}
\acrodef{ANN}{Artificial Neural Network}
\acrodef{UE}{User Equipment}
\acrodef{BS}{Base Station}
\acrodef{DTM}{Digital Terrain Model}
\acrodef{DHM}{Digital Height Model}
\acrodef{DLU}{Digital Land Use Map}
\acrodef{LoS}{Line of Sight}
\acrodef{NLoS}{Non Line of Sight}
\acrodef{k-NN}{k-Nearest Neighbors}
\acrodef{DT}{Decision Tree}
\acrodef{RMSE}{Root Mean Square Error}
\acrodef{GBDT}{Gradient Boosting Decision Trees}
\acrodef{AdaBoost}{Adaptive Boosting}
\acrodef{XGBoost}{Extreme Gradient Boosting}
\acrodef{LightGBM}{Light Gradient Boosting Machine}
\acrodef{CatBoost}{Categorical Boosting}
\acrodef{DNN}{Deep Neural Network}
\acrodef{LIME}{Local Interpretable Model-Agnostic Explanations}
\acrodef{SHAP}{SHapley Additive exPlanations}
\acrodef{MDT}{Minimization of Drive Tests}
\acrodef{$R^2$}{coefficient of determination}
\acrodef{URLLC}{Ultra-Reliable Low-Latency Communication}
\acrodef{UAV}{Unmanned Aerial Vehicle}
\acrodef{RIS}{Reconfigurable Intelligent Surfaces}

\begin{document}

\title{Interpretable AI-based Large-scale 3D Pathloss Prediction Model for enabling Emerging Self-Driving Networks}

\author{Usama~Masood,~\IEEEmembership{Graduate Student Member,~IEEE,}
        Hasan~Farooq,~\IEEEmembership{Member,~IEEE,}
        Ali~Imran,~\IEEEmembership{Senior~Member,~IEEE}
        and~Adnan~Abu-Dayya,~\IEEEmembership{Senior~Member,~IEEE}
\IEEEcompsocitemizethanks{\IEEEcompsocthanksitem
This work was supported in part by the National Science Foundation (NSF) under Grant 1559483, Grant 1619346, and Grant 1730650; and in part by the Qatar National Research Fund (QNRF) under Grant NPRP12-S 0311-190302.
}
\IEEEcompsocitemizethanks{\IEEEcompsocthanksitem\vspace{-6pt}
A portion of this work has been published in IEEE GlobeCom'19 \cite{Maso1912:Machine}.
}
\IEEEcompsocitemizethanks{\IEEEcompsocthanksitem\vspace{-6pt}
Usama~Masood and Ali~Imran are with the AI4Networks Research Center, School of Electrical \& Computer Engineering, University of Oklahoma, Tulsa, OK, 74135 USA. E-mail: \{usama.masood, ali.imran\}@ou.edu
}
\IEEEcompsocitemizethanks{\IEEEcompsocthanksitem\vspace{-6pt}
Hasan~Farooq is with the Ericsson Research, USA. E-mail: hasan.farooq@ericsson.com
}
\IEEEcompsocitemizethanks{\IEEEcompsocthanksitem\vspace{-6pt}
Adnan~Abu-Dayya is with the Electrical Engineering Department, Qatar University. E-mail: adnan@qu.edu.qa
}
}
\markboth{Accepted at IEEE Transactions on Mobile Computing}%
{Masood \MakeLowercase{\textit{et al.}}: Interpretable AI-based Large-scale 3D Pathloss Prediction Model for enabling Emerging Self-Driving Networks}

\IEEEtitleabstractindextext{

\begin{abstract}
In modern wireless communication systems, radio propagation modeling to estimate pathloss has always been a fundamental task in system design and optimization. The state-of-the-art empirical propagation models are based on measurements in specific environments and limited in their ability to capture idiosyncrasies of various propagation environments. To cope with this problem, ray-tracing based solutions are used in commercial planning tools, but they tend to be extremely time-consuming and expensive. We propose a \acf{ML}-based model that leverages novel key predictors for estimating pathloss. By quantitatively evaluating the ability of various ML algorithms in terms of predictive, generalization and computational performance, our results show that \acf{LightGBM} algorithm overall outperforms others, even with sparse training data, by providing a 65\% increase in prediction accuracy as compared to empirical models and 13x decrease in prediction time as compared to ray-tracing. To address the interpretability challenge that thwarts the adoption of most \ac{ML}-based models, we perform extensive secondary analysis using \acf{SHAP} method, yielding many practically useful insights that can be leveraged for intelligently tuning the network configuration, selective enrichment of training data in real networks and for building lighter \ac{ML}-based propagation model to enable low-latency use-cases.

\end{abstract}


}

\maketitle

\IEEEdisplaynontitleabstractindextext

\IEEEpeerreviewmaketitle

\acresetall

\IEEEraisesectionheading{\section{Introduction}\label{sec:introduction}}

\IEEEPARstart{E}{merging} cellular networks are anticipated to witness a dramatic growth in connected devices and exciting new vertical services. \ac{AI} enabled end to end network automation vis-a-vis next generation Self Organizing Network (AISON), as proposed in \cite{imran2014challenges}, is considered to be the key enabler to meet the stringent performance requirements of increasingly complex planning, operation, optimization and maintenance in emerging self-driving networks. 

The ultimate goal of AISON is to autonomously orchestrate the plethora of network parameters to maintain optimal multi-faceted network performance in terms of all important top level \acp{KPI}, without much human involvement \cite{imran2014challenges}. Most of the high level KPIs such as capacity, \ac{QoS} and energy efficiency ultimately depend on one core low level metric i.e., \ac{RSS}. Therefore, characterising RSS as a function of network parameters is the very first step towards optimally designing and operating a cellular network. Hence, a realistic propagation model that is sensitive to the variations in network parameters (e.g., tilt) and environment geography and can follow the spatio-temporal variation in the network will be the cornerstone of AISON enabled future cellular networks (5G and beyond).

The existing propagation models can be categorized into three classes: deterministic, empirical and semi-empirical \cite{sarkar2003survey, phillips2012survey}; \textit{deterministic models} are based on the principles of wave propagation that can be theoretically computed using Maxwell's equations. However, in practice approximate methods such as \textit{ray tracing} are used to model signal propagation, by taking into account the interactions of rays with the  environment and using the dominant ray path to calculate the pathloss. These models can be very accurate depending on the resolution of available topographical database, but unfortunately are \textit{computationally inefficient}. On the other hand, \textit{empirical} and \textit{semi-empirical models} such as COST-Hata \cite{1991urban}, \ac{SUI} \cite{erceg2001channel}, \ac{SPM} \cite{Atoll:MCG} and ITU-R P.452-15 \cite{ITU452} can be efficiently computed. However, these empirical and semi-empirical models are less sensitive to the actual physical and geometric structure of a given propagation environment and require in-depth domain-specific knowledge and technical expertise in radio signal propagation across electromagnetic fields.

To address the constraints and limitations of traditional propagation modeling methods, Artificial Intelligence and \ac{ML} techniques are being considered as promising viable solutions and have been proven to be very effective for approximating arbitrary functions with hidden features. As envisioned in \cite{imran2014challenges}, AI is going to be indispensable in optimally designing and operating increasingly complex cellular networks. Hence, \ac{AI} can replace classical mathematical models with a robust data-driven pathloss prediction model, that is more accurate than empirical propagation models and more computationally efficient than deterministic models, for system level intelligent network planning and post-deployment optimization and automation in cellular networks.

\subsection{Related Work}
In recent years several studies have been conducted for pathloss prediction in a particular environment using machine learning based models. \acp{ANN} have been at the core of most \ac{ML}-based pathloss prediction models, however, the input features to the \ac{ANN} model in these studies are limited to a particular environment, such as rural \cite{ostlin2010macrocell}, urban \cite{piacentini2011path,zhang2019path} and volcanic ocean islands \cite{ribero2019deep}, and seems unable to scale to other environment settings. The authors in \cite{sotiroudis2013application} went one step ahead and used evolutionary algorithms to find the optimal hyper-parameters of the ANN based model, but they assumed a uniformly structured simulation area, which is not the case in practical scenarios. The authors in \cite{ayadi2017uhf} incorporated features based on clutter maps to differentiate between different environments, but the presented model is still unable to capture the variation in coverage due to the change in geometrical structure of the propagation path. Furthermore, the authors in \cite{popoola2018optimal} tried to capture this variation by incorporating clutter heights as input feature, but still their input feature set is very limited to scale to different network configuration, as is the case in real networks. On the other hand, supervised \ac{ML} techniques have also been used for pathloss prediction. The authors in \cite{timoteo2014proposal} used Support Vector Regression to predict pathloss. However, they trained their model on a drive test data from a single serving \ac{BS} in an urban environment. The authors in \cite{dai2018propagation} compared the performance of several supervised ML algorithms for estimating cellular network coverage, using \ac{UE} measurement traces, \ac{BS} parameters and geographical information. However, instead of modeling the pathloss or RSS, the authors classify the observation area as a good or a bad coverage area, using a pre-defined coverage threshold. A recently proposed data-driven model in \cite{ghasemi2018data} is the most relevant to our framework, using a boosting ensemble learning method to predict \ac{RSS} using \ac{UE} data from crowdsourcing applications. However, the environmental features used in the model are very limited.

\begin{figure*}
    \centering
    \includegraphics[trim={1cm 1cm 1cm 1cm}, clip, width=0.99\linewidth]{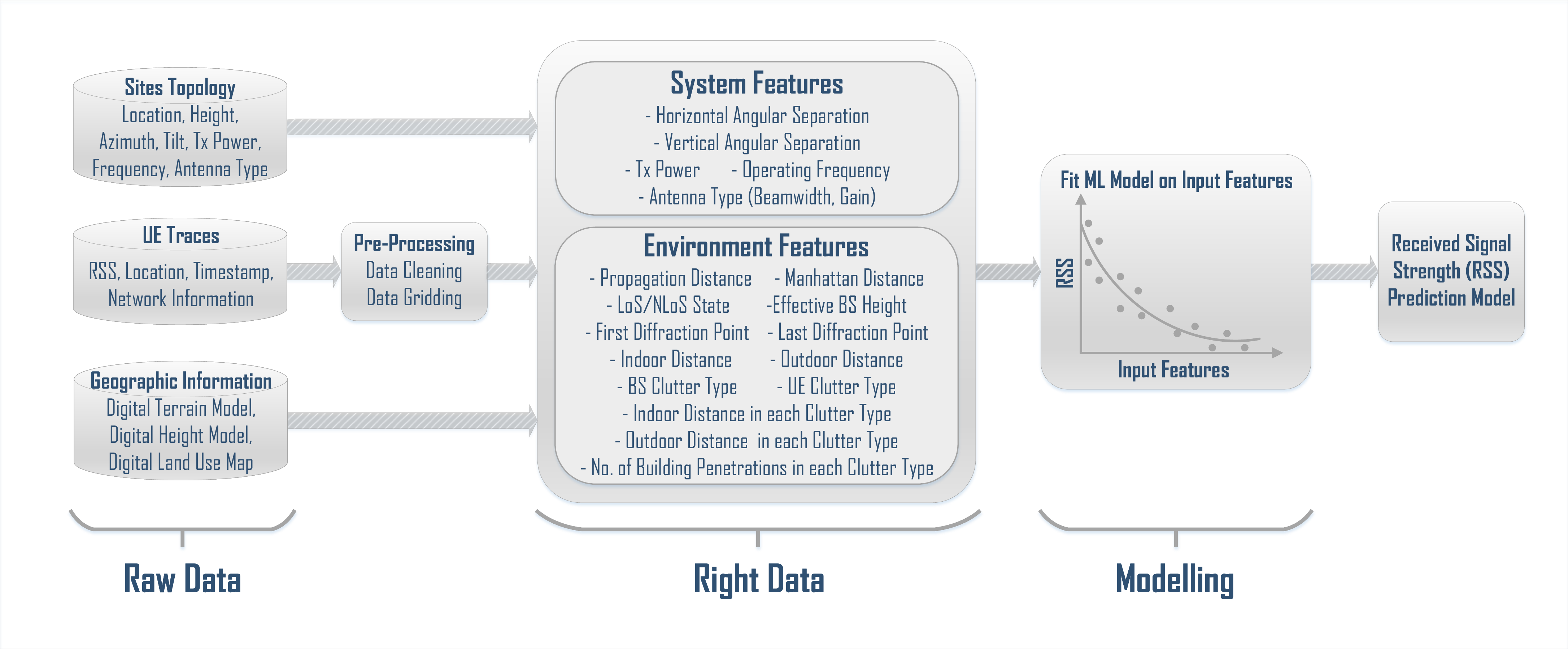}
    \caption{Proposed Framework of a Machine Learning-based 3D Radio Propagation Model. Raw datasets such as BS sites topology, UE measurement traces and geographical information are first pre-processed and then converted into right data i-e- feature matrix comprising of key features/predictors that can estimate pathloss (or RSS). These key features are then used to train state-of-the-art ML algorithms for creating an RSS prediction model.}
    \label{fig:system_model}
\end{figure*}

\subsection{Contributions and Organization}
In this paper, we present a framework for an \ac{AI}-driven large-scale 3D pathloss model (See \cref{fig:system_model}), that is scalable and robust to the variations in the environment geography and addresses the limitations of aforementioned studies. The contributions of this paper can be summarized as follows:
\begin{enumerate}
       \item A novel set of key predictors (features) are identified that can characterize the physical and geometric structure of the environment traversed by a signal in its propagation path (e.g., \textit{indoor distance, Manhattan distance, number of building penetrations in each clutter type}).
      \vspace{2pt}
       \item Multi-faceted performance comparison of the current state-of-the-art ML algorithms is done to evaluate the ability of proposed ML-based propagation model for real-time implementation, that includes predictive performance, generalization performance (robustness to unseen propagation scenarios using sparse training data, as is the case in real networks) and computational performance (i.e., training time and prediction time). In our investigation, \acf{LightGBM} algorithm is found to be the most optimal choice overall in modeling the complex propagation environment in real networks, due to its lightning fast training process and robustness to sparsity of training data.
      \vspace{2pt}
       \item The baseline performance of LightGBM algorithm is further optimized by investigating four different hyperparameter optimization approaches. These include Grid Search, Random Search, Bayesian Optimization and Simulated Annealing. These approaches are investigated in terms of performance gain and computational complexity (convergence time), and Bayesian optimization seems to be the best approach among them, as it converges in just 3 search iterations and reduces the prediction RMSE by 10\%.
      \vspace{2pt}
       \item Performance comparison of the proposed model with state-of-the-art empirical propagation models and ray-tracing approach is also provided. The results show a 65\% increase in prediction accuracy as compared to empirical propagation models and 13x decrease in prediction time as compared to ray tracing.
      \vspace{2pt}       
       \item A key caveat of using \ac{ML} is the lack of interpretability of resultant models, i.e., the black box paradox. In this study we try to address this weakness of the proposed \ac{RSS} estimation models by performing extensive secondary analysis of the proposed models through SHAP method to interpret the model's predictions and bring clarity in understanding the importance of each feature (e.g., Azimuth, Tilt and Antenna Height) in the model. Utility of the insights drawn from the secondary SHAP analysis are also provided, such as intelligent optimization of network configuration, smart/selective enrichment of training data in real networks and building lighter ML-based propagation model to enable low-latency use-cases.
      \vspace{2pt}
       \item Another key contribution of the paper is to leverage the SHAP interpretability analysis to improve the various aspects of performance in the baseline model. We leverage the insights from SHAP analysis to propose a second novel light weight model for real-time implementation. This second model uses only the most significant features (selected based on insights gained from SHAP analysis) and as a result has $\sim$ 70\% less computational complexity compared to the base line model at the cost of negligible loss in performance (around 3\%).
       \vspace{2pt}
\end{enumerate}

The rest of the paper is organized as follows: \cref{section2} explains the proposed framework, starting from raw data from the network, data pre-processing, feature engineering and a comparison of various machine learning algorithms for modeling \ac{RSS}, whereas, \cref{section3} provides the performance comparison of the proposed model with traditional propagation models. Interpretation of the proposed \ac{AI}-driven model using a recently proposed sensitivity analysis technique is given in \cref{section4} and finally \cref{section5} concludes this paper.

\section{Proposed Framework}\label{section2}
The proposed framework for an \ac{AI}-driven 3D propagation model (\cref{fig:system_model}) uses raw data from the network consisting of network topology information, \ac{UE} measurement traces and geographic information of the area, pre-process them and converts them into right data \cite{imran2014challenges}, which is then fed to a \ac{ML} model to estimate \ac{RSS} at given locations in a radio propagation environment.

\subsection{Network and Simulation Setup}

\begin{table}[!h]
        \begin{center}
                \caption{Network Scenario Settings}
                \label{tab:simulation}
                \renewcommand{\arraystretch}{1.1}
                \vspace{-5pt}
                \begin{tabular}{|c|c|}
                        \hline
                        \thead{System Parameters} & \thead{Values} \\
                        \hline
                        Air Technology & 4G LTE \\
                        \hline
                        Cellular Layout & 10 Macrocell sites \\
                        \hline
                        Sectors & 3 sectors per eNB\\
                        \hline
                        Simulation Location & Brussels, Belgium\\
                        \hline
                        Simulation Area & 3.80 km$^2$\\
                        \hline
                        User Distribution & Poisson Distribution\\
                        \hline
                        Propagation Model & Aster (Ray Tracing)\\
                        \hline
                        Path Loss Matrix Resolution & 10m\\
                        \hline
                        Geographic Information & \vtop{\hbox{\strut (\textit{1-m Resolution GeoData})}\hbox{\strut Ground Heights (DTM) + }\hbox{\strut Building Heights (DHM) + }\hbox{\strut Land Use Map (DLU)}}\\
                        \hline
                        Land Cover (Clutter) Types & 15 different classes\\
                        \hline
                        eNB Transmit Power (max) & 43 dBm\\
                        \hline
                        eNB Noise Figure & 5 dB\\
                        \hline
                        eNB Antenna Height & Actual site heights\\
                        \hline
                        eNB Antenna Model & \vtop{\hbox{\strut Kathrein Directional Antenna}\hbox{\strut (Model 742 265)}}\\
                        \hline
                        eNB Antenna Gain & 18.3 dBi\\
                        \hline
                        \vtop{\hbox{\strut eNB Antenna Horizontal}\hbox{\strut Half Power Beamwidth}} & 63$^o$\\
                        \hline
                        \vtop{\hbox{\strut eNB Antenna Vertical}\hbox{\strut Half Power Beamwidth}} & 4.7$^o$\\
                        \hline
                        Frequency Band & 2110 FDD (E-UTRA Band 1)\\
                        \hline
                        Channel Bandwidth (CBW) & 5 MHz\\
                        \hline
                \end{tabular}
        \end{center}
        
\end{table}

\begin{figure*}[!ht]
        \centering     
        \subfigure[3D Elevation Map]{\includegraphics[width=.39\linewidth]{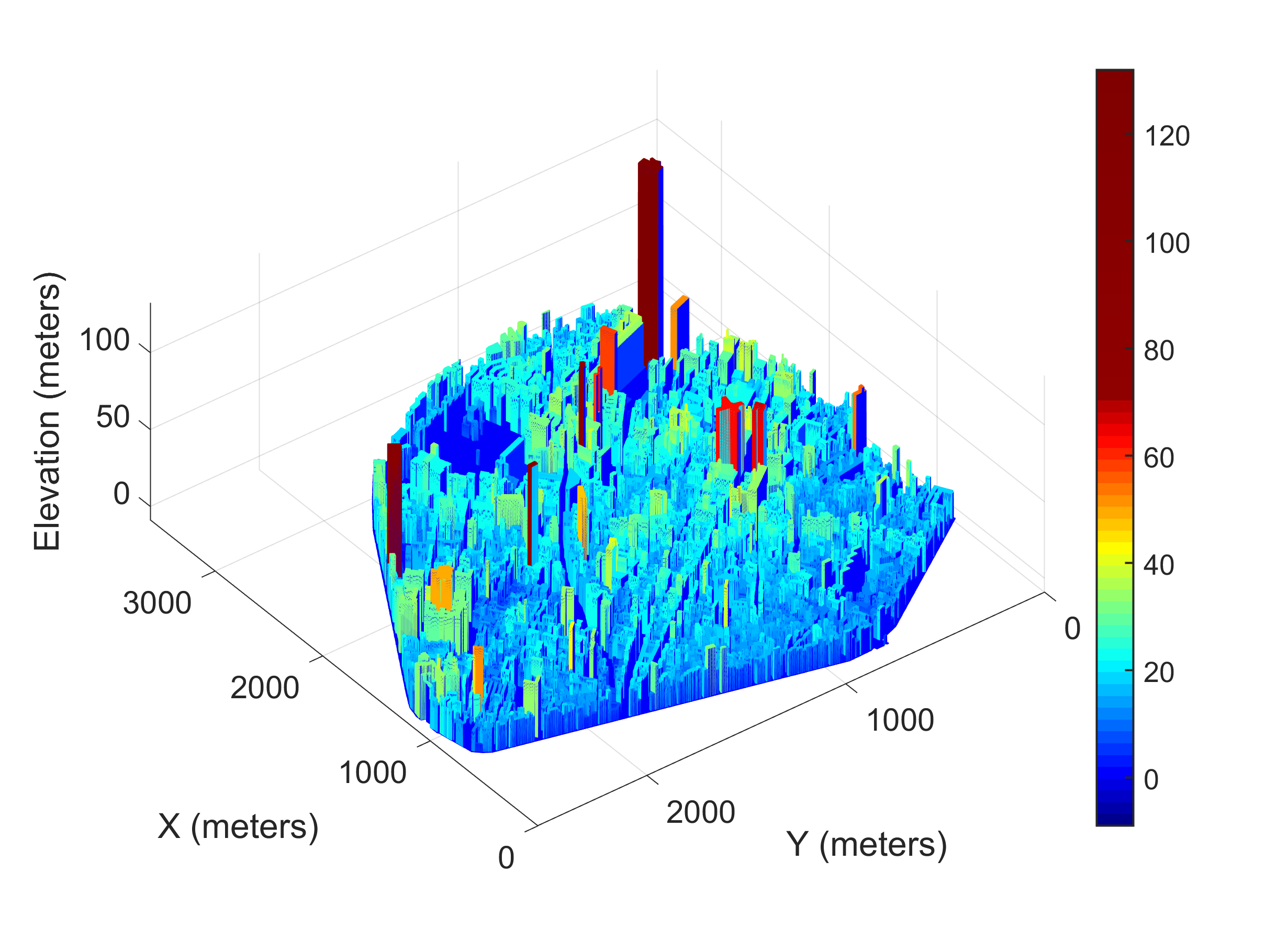}\label{fig:simulation1}}
        \subfigure[2D Elevation Map]{\scalebox{-1}[1]{{\includegraphics[width=.24\linewidth]{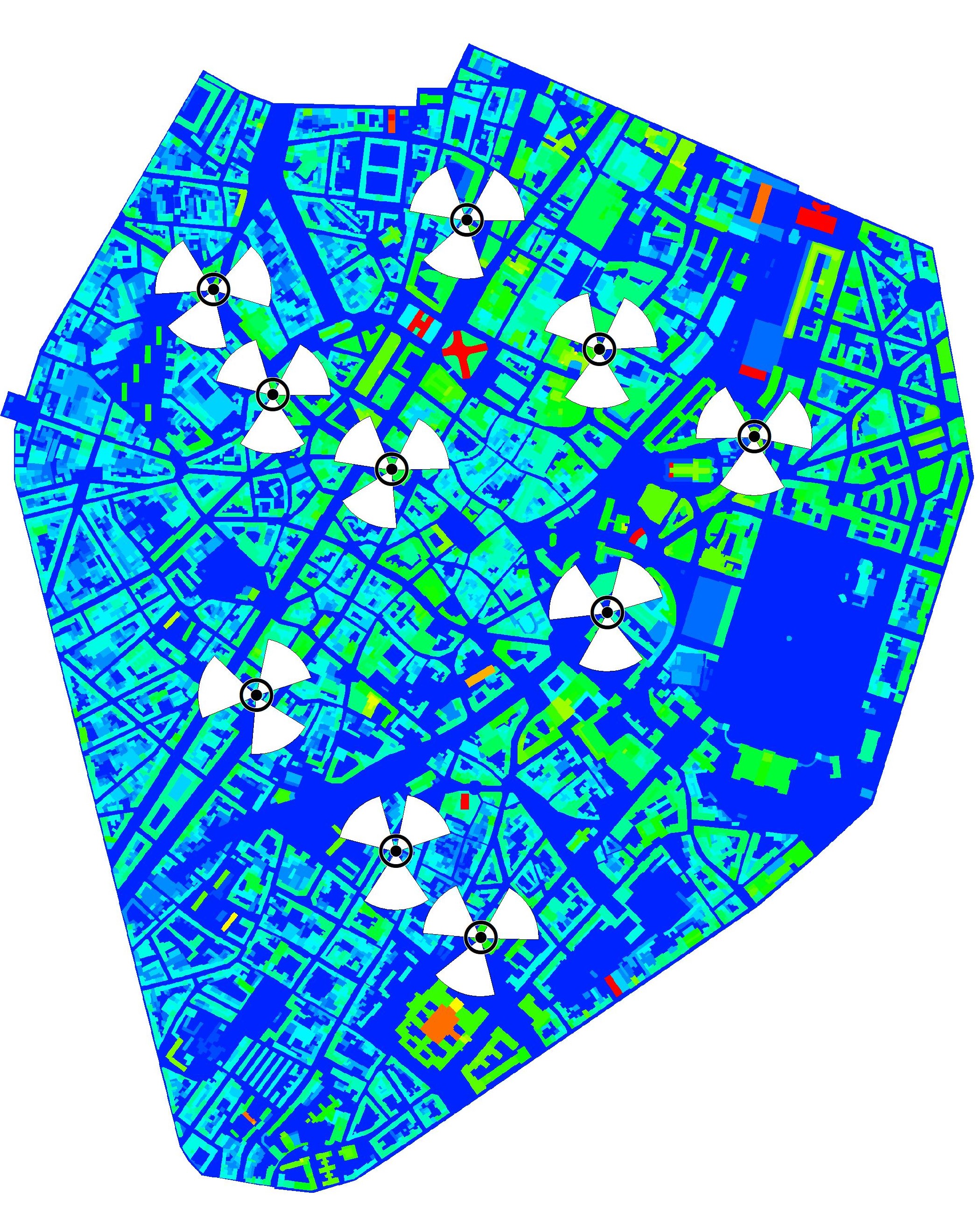}\label{fig:simulation2}}}}
        \subfigure[Propagation Path between a BS and UE]{
        \includegraphics[width=.32\linewidth]{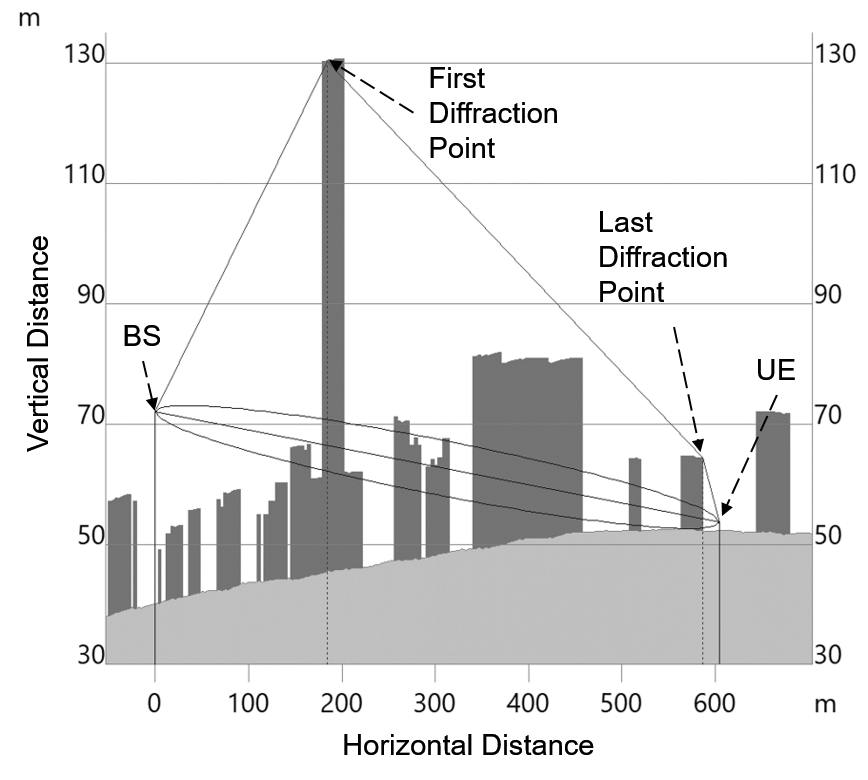}\label{fig:simulation3}}
        \caption{Area of Simulation showing ({\bf a}) Building Heights ({\bf b}) Transmitter Positions and ({\bf c}) Vertical Propagation Path}
        \label{fig:simulation}
\end{figure*}

\begin{itemize}

\item \uline{Network Scenario:}
A ray-tracing based industry standard radio network planning and optimization platform ``Atoll'' \cite{Atoll} is used to create a sophisticated network topology, consisting of 10 macro cell sites  in the center of City of Brussels, Belgium (See \cref{fig:simulation2}). Actual antenna pattern and antenna heights are used in our analysis. \cref{tab:simulation} lists all the network and simulation scenario settings used in our analysis. 

\vspace{5pt}
\item \uline{Geographical Datasets:} High resolution (1-m) geographical datasets containing earth terrain, buildings heights and land use information (e.g., Open, Parks, Block Buildings etc.) of the city are used in our analysis, enabling realistic and accurate pathloss calculation using Aster high-performance propagation model \cite{Aster}.

\item \uline{Ray Tracing:} Aster utilizes advanced ray-tracing propagation techniques to calculate various phenomena that affects radio wave propagation including vertical diffractions over roof-tops, horizontal diffractions and reflections based on ray-launching, atmospheric absorptions, rain attenuation and vegetation through loss etc. Aster utilizes advanced ray-tracing propagation techniques to calculate various phenomena that affects radio wave propagation including vertical diffractions over roof-tops, horizontal diffractions and reflections based on ray-launching, atmospheric absorptions, rain attenuation and vegetation through loss etc. 

\vspace{5pt}
\item \uline{Validation using Real Data:} Furthermore, the parameters of the Aster propagation model are pre-calibrated using more than 1.5 million real channel measurement points from the real environment \cite{Aster:TRG}. Therefore, the high-fidelity \ac{RSS} (or pathloss) data calculated by Atoll in the observation area can be taken as ground truth for designing a realistic propagation model.

\end{itemize}

\subsection{Raw Data}
The following three different kinds of datasets are required as input data for our proposed framework:

\begin{enumerate}
\item \uline{Sites Topology:} This dataset contains the Location, Height, Azimuth, Tilt, Transmit Power, Frequency, Antenna Type of all the \acp{BS} in the observation area. It is denoted by ${D}_{\textit{BS}}$ (See \cref{tab:bs}).

\begin{table}[!h]
        \begin{center}
                \caption{Sites Topology}
                \label{tab:bs}
                \renewcommand{\arraystretch}{1.3}
                \vspace{-5pt}
                \begin{tabular}{|l|l|}
                        \hline
                        \thead{Parameter} & \thead{Description}\\
                        \hline
                        \vtop{\hbox{\strut Location}\hbox{\strut ($x_{\textit{BS}},y_{\textit{BS}}$)}} & Location coordinates of a BS site (See \cref{fig:simulation2})\\
                        \hline
                        \vtop{\hbox{\strut Height}\hbox{\strut ($h_{\textit{BS}}$)}} & \vtop{\hbox{\strut The height of BS antenna above the ground}\hbox{\strut and building (if any)}} \\
                        \hline
                        \vtop{\hbox{\strut Azimuth}\hbox{\strut ($\theta_{\textit{BS}}$)}} & \vtop{\hbox{\strut Azimuth angle (in degrees) of the BS antenna,}\hbox{\strut which is the direction of antenna w.r.t. North}}\\
                        \hline
                        \vtop{\hbox{\strut Tilt}\hbox{\strut ($\phi_{\textit{BS}}$)}} & \vtop{\hbox{\strut Tilt angle (in degrees) of the BS antenna, which}\hbox{\strut is basically the angle below the horizontal plane}}\\
                        \hline
                        \vtop{\hbox{\strut Tx Power}\hbox{\strut ($P_{\textit{BS}}$)}} & \vtop{\hbox{\strut The power of the radio signal (in dBm) when} \hbox{\strut it's transmitted from the BS antenna}}\\
                        \hline
                        \vtop{\hbox{\strut Frequency}\hbox{\strut ($f$)}} & \vtop{\hbox{\strut Carrier frequency used by the BS antenna for}\hbox{\strut transmission}}\\
                        \hline
                        \vtop{\hbox{\strut Antenna}\hbox{\strut Type}} & \vtop{\hbox{\strut The type of antenna used by the BS transmitter.}\hbox{\strut It is differentiated by beamwidth, antenna gain etc}}\\
                        \hline
                \end{tabular}
        \end{center}
\end{table}

\item \uline{Geographic Information:} The geographical information of the propagation environment can be captured using three types of geographical datasets: \ac{DTM}, \ac{DHM} and \ac{DLU}. These datasets are in a raster grid format, which means that the whole observation area is divided into grids (or bins), each grid containing a specific value (See \cref{tab:geo}). These geographical datasets are routinely used by mobile telecom industry for their planning and maintenance tasks, and can be acquired on demand \cite{EGS}.

\begin{table}[!h]
        \begin{center}
                \caption{Geographic Information}
                \label{tab:geo}
                \renewcommand{\arraystretch}{1.3}
                \vspace{-5pt}
                \begin{tabular}{|l|l|}
                        \hline
                        \thead{Parameter} & \thead{Description}\\
                        \hline
                        \vtop{\hbox{\strut Digital Terrain}\hbox{\strut Model ($D_{\textit{DTM}}$)}} & \vtop{\hbox{\strut It provides the earth terrain (ground) height.}\hbox{\strut  It takes as an input the $x,y$ coordinates and}\hbox{\strut outputs the ground height $z_{\textit{DTM}}$ at that place}\hbox{\strut  $z_{\textit{DTM}}=D_{\textit{DTM}}(x,y)$. (See \cref{fig:simulation3})}}\\
                        \hline
                        \vtop{\hbox{\strut Digital Height}\hbox{\strut Model ($D_{\textit{DHM}}$)}} & \vtop{\hbox{\strut It provides the building heights (above the}\hbox{\strut  ground) in the observation area. It takes}\hbox{\strut as an input the $x,y$ coordinates and output}\hbox{\strut the total height $z_{\textit{DHM}}$ of the user at that place}\hbox{\strut  $z_{\textit{DHM}}=D_{\textit{DHM}}(x,y)$. (See \cref{fig:simulation1})}}\\
                        \hline
                        \vtop{\hbox{\strut Digital Land}\hbox{\strut Use Map ($D_{\textit{DLU}}$)}} & \vtop{\hbox{\strut It provides the clutter (or land cover)}\hbox{\strut  type of each grid in the observation area.}\hbox{\strut It takes as an input the $x,y$ coordinates}\hbox{\strut and output the clutter type $c$ at that place}\hbox{\strut  $c=D_{\textit{DLU}}(x,y)$}}\\
                        \hline
                \end{tabular}
        \end{center}
\end{table}

\item \uline{UE Traces:} This dataset contains the RSS, Location, Timestamp, Network ID of all the UEs in the observation area. It is denoted by ${D}_{\textit{UE}}$ (See \cref{tab:ue}). The mobile operators can readily use the data from Drive Tests, \ac{MDT} reports, crowdsourcing applications etc. to generate this dataset, without the need for any new standardization. 

\end{enumerate}
\begin{table}[!h]
        \begin{center}
                \caption{UE Traces}
                \label{tab:ue}
                \renewcommand{\arraystretch}{1.3}
                \vspace{-5pt}
                \begin{tabular}{|l|l|}
                        \hline
                        \thead{Parameter} & \thead{Description}\\
                        \hline
                        RSS & \vtop{\hbox{\strut Received Signal Strength ($P_{\textit{UE}}$) from the serving}\hbox{\strut Base Station (BS)}}\\
                        \hline
                        Location & \vtop{\hbox{\strut Location coordinates ($x_{\textit{UE}},y_{\textit{UE}}$) of a UE}}\\
                        \hline
                        Timestamp & \vtop{\hbox{\strut Time at which the UE trace is recorded}}\\
                        \hline
                        Network ID & \vtop{\hbox{\strut Information regarding serving BS ID,}\hbox{\strut Mobile Network Code etc.}}\\
                        \hline
                \end{tabular}
        \end{center}
\end{table}

\subsection{Data Pre-processing}
Raw UE traces from the network are first pre-processed by cleaning and gridding, before using them for feature extraction.

\subsubsection{Data cleaning}
Data cleaning is the process of identifying missing and corrupt values in the dataset and then handling them by modifying or deleting the relevant rows (or entries) from the dataset. This pre-processing step ensures that the training data for the proposed model is free from any anomalies and inconsistencies. In our study, some UE traces containing missing values (e.g., out of coverage UEs) are removed before using them for further analysis. 

\subsubsection{Data gridding}

Data gridding is the process of mapping all \ac{UE} traces into unique spatial bins (of 10m width in our case) and then averaging the measurements inside each spatial bin for every serving BS. The advantage of data gridding is twofold:
\begin{enumerate}
    \item \uline{Handling Randomness:} Firstly, it can offset randomness in \ac{RSS} due to fast fading and slow fading (to some extent), by averaging all measurements from the same BS, falling within a small bin, given the \ac{RSS} within the bin is expected to stay almost same due to its small size.
    \item \uline{Handling Positioning Error:} Secondly, it handles positioning error in the user reported measurements. However, gridding/binning has its costs i.e., it introduces quantization error to say the least and also presents an accuracy vs complexity trade-off.
\end{enumerate}

In our study, UE traces in the observation area are first mapped into unique spatial bins of 10m width, and then in each bin all UE traces corresponding to a unique BS were averaged out. For further analysis, the averaged UE traces are used to mitigate the effect of randomness from the original data.

For detailed analysis of the impact of gridding, reader is referred to a recent study in \cite{qureshi2019optimal} and \cite{qureshi2020enhanced}. Analysis presented in \cite{qureshi2019optimal} and \cite{qureshi2020enhanced} shows that there exists a trade-off between the quantization error introduced by gridding and the positioning error from the incorrect GPS location tagged with the \ac{UE} measurements, and that there exists an optimal bin-width for gridding for a given \ac{UE} density and positioning error that maximizes the accuracy of \ac{UE} measurements data.

\begin{figure*}[!ht]
        \centering
        \includegraphics[trim={0.5cm 0.5cm 0.5cm 0.5cm}, clip, width=0.89\linewidth]{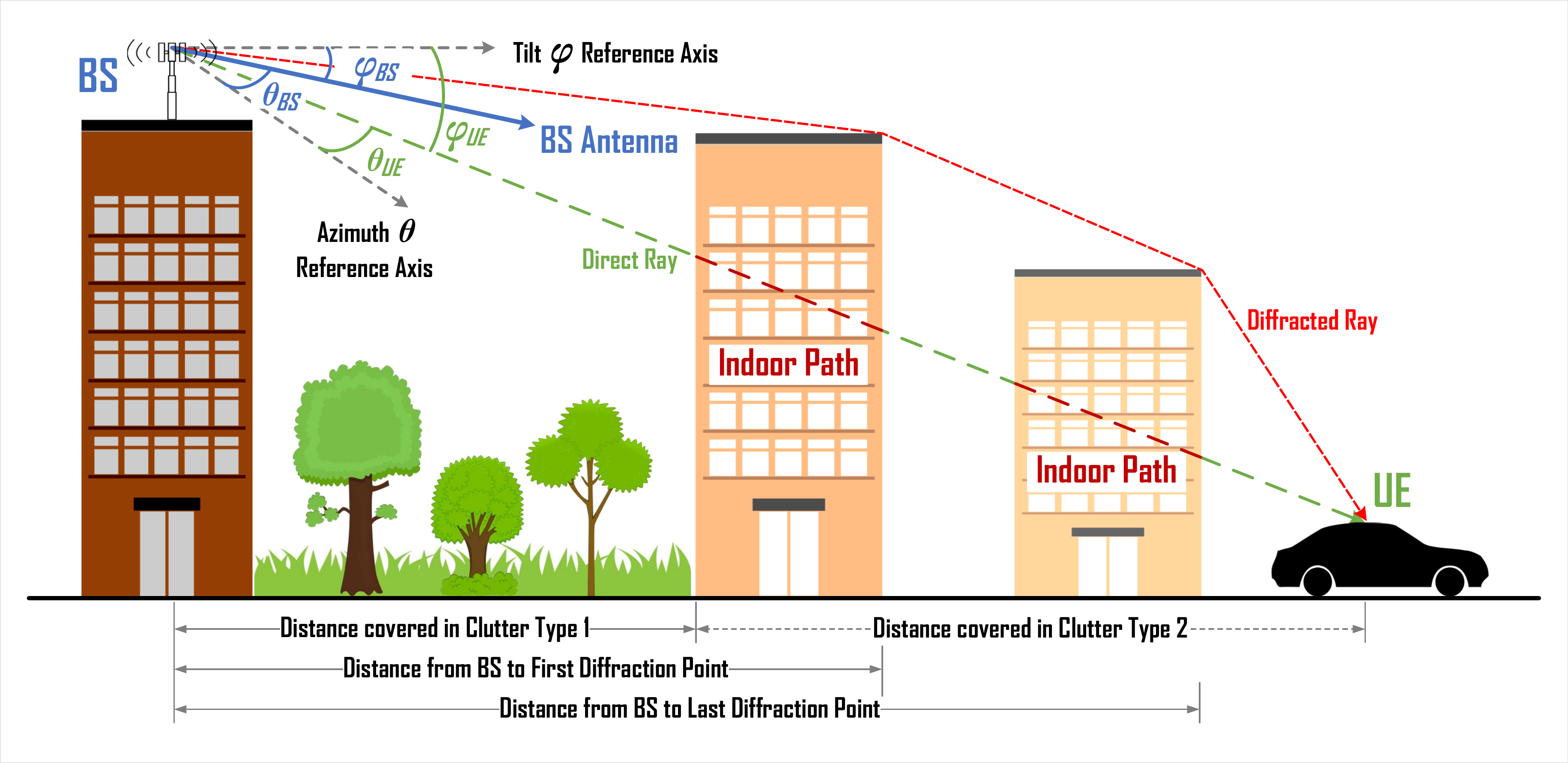}
        \caption{Propagation Path between a BS and UE, showing various features of the proposed model}
        \label{fig:key_features}
\end{figure*}

\subsection{Feature Engineering (Raw Data to Right Data)}
Feature engineering is a key process in ML, that leverages domain knowledge to create features which can characterize the complex target model and greatly enhance its learning performance. 

In our study, several \textit{key predictors (features)} are identified or engineered, to better characterize the environment traversed by a signal in its propagation path (See \cref{fig:key_features}). The raw network, UE and geographic datasets, readily available to the mobile operators, are converted into right data (key features) comprising of system as well as propagation environment features that can then be leveraged to train an ML-based propagation model.

\subsubsection{Propagation Distance}
This is the horizontal distance (in meters) between a UE and its serving BS. It is denoted by $d$.
\begin{equation}
\label{eq:stress}
d=\sqrt{(x_{\textit{BS}}-x_{\textit{UE}})^2+(y_{\textit{BS}}-y_{\textit{UE}})^2}.
\end{equation}

\subsubsection{Horizontal Angular Separation}
This is the horizontal angular separation (in degrees) between the BS antenna boresight and the direction of Line of Sight path to the UE. This feature captures the attenuation due to horizontal antenna pattern of the BS. It is denoted by $\theta_{hor}$.
\begin{equation}
\label{eq:hor1}
\theta_{\textit{hor}}=abs(\theta_{\textit{BS}}-\theta_{\textit{UE}}),
\end{equation}
where,
\begin{equation}
\label{eq:hor2}
\theta_{\textit{UE}}=atan2\left(\frac{x_{\textit{BS}}-x_{\textit{UE}}}{y_{\textit{BS}}-y_{\textit{UE}}}\right).
\end{equation}
Here $\theta_{\textit{UE}}$ is the azimuth angle of (UE) arrival and $\theta_{\textit{BS}}$ is the azimuth angle of (BS) departure, or simply BS azimuth angle, whereas, $atan2()$ calculates the four quadrant inverse tangent.

\subsubsection{Vertical Angular Separation}
This is the vertical angular separation (in degrees) between the BS antenna boresight and the direction of Line of Sight path to the UE. This feature captures the attenuation due to vertical antenna pattern of the BS. It is denoted by $\phi_{ver}$. 
\begin{equation}
\label{eq:ver1}
\phi_{\textit{ver}}=\phi_{\textit{UE}}-\phi_{\textit{BS}},
\end{equation}
where,
\begin{equation}
\label{eq:ver2}
\phi_{\textit{UE}}=atan\left(\frac{z_{\textit{UE}}-z_{\textit{BS}}}{d}\right),
\end{equation}
\begin{equation}
\label{eq:ver3}
z_{\textit{UE}}=D_{\textit{DTM}}(x_{\textit{UE}},y_{\textit{UE}})+D_{\textit{DHM}}(x_{\textit{UE}},y_{\textit{UE}}),
\end{equation}

\begin{equation}
\label{eq:ver4}
z_{\textit{BS}}=D_{\textit{DTM}}(x_{\textit{BS}},y_{\textit{BS}})+D_{\textit{DHM}}(x_{\textit{BS}},y_{\textit{BS}}).
\end{equation}

Here $\phi_{\textit{UE}}$ is the tilt angle of (UE) arrival, $\phi_{\textit{BS}}$ is the tilt angle of (BS) departure, or simply BS tilt angle, $z_{\textit{UE}}$ is the total height of UE and $z_{\textit{BS}}$ is the total height of BS (See \cref{tab:geo} for details on $D_{\textit{DTM}}$ and $D_{\textit{DHM}}$).

\subsubsection{Effective BS Height}
This is the vertical distance (in meters) between a UE and its serving BS. It is denoted by $d_{\textit{vert}}$.
\begin{equation}
\label{eq:bs_height}
d_{\textit{vert}}=z_{\textit{BS}} - z_{\textit{UE}}.
\end{equation}

\subsubsection{Manhattan Distance}
This represents the Manhattan distance between the \ac{BS} and \ac{UE}. As radio waves also diffracts at the street corners and are more likely to travel along the streets in urban areas, therefore Manhattan distance is a better metric to calculate the distance traversed by the signal, especially in urban networks \cite{karttunen2017spatially, wang2018mmwave}. It is denoted by $d_{\textit{man}}$.

\subsubsection{LoS / NLoS State}
This represents the link status between a BS and a UE antenna. A UE can either be in a \ac{LoS} or \ac{NLoS} region from the \ac{BS}. This feature is particularly useful in wireless channels, as \ac{LoS} regions have higher \ac{RSS}, and vice versa. It is denoted by $L$.

\subsubsection{First Diffraction Point}
This is the horizontal distance (in meters) from a BS to the first diffraction point in the propagation path between a BS and UE. This feature captures the significance of diffracted rays at the receiver, as multiple rays from the same BS are received and the ray having the highest signal strength is selected as the dominant ray. It is denoted by $d_{\textit{FD}}$.

\subsubsection{Last Diffraction Point}
This is the horizontal distance (in meters) from a BS to the last diffraction point in the propagation path between a BS and UE. This feature also tries to learn the behavior of diffracted rays in the estimation of RSS. It is denoted by $d_{\textit{LD}}$.

\subsubsection{Number of Building Penetrations} This is the number of buildings penetrated by the signal in its direct path between a BS and UE. This feature characterizes the penetration loss (dB) experienced by the signal while crossing buildings. It is denoted by $N$.

\subsubsection{Indoor Distance}
Horizontal distance (in meters) in the direct path between a BS and UE that is passing through buildings (indoor). This feature characterizes the linear loss (dBm/m) experienced by the signal in an indoor area. It is denoted by $d_{\textit{indoor}}$.

\subsubsection{Outdoor Distance}
Horizontal distance (in meters) in the direct path between a BS and UE that is in open area (outdoor). This feature characterizes the linear loss (dBm/m) experienced by the signal in an open area. It is denoted by $d_{\textit{outdoor}}$.

\subsubsection{BS Clutter Type}
It is the clutter type (or land cover type) of the \ac{BS} (For example: open street, dense buildings, sparse buildings, trees, water etc.). This feature tries to learn the effect of different land cover type on the signal around the \ac{BS} antenna. It is denoted by $c_{\textit{BS}}$.

\subsubsection{UE Clutter Type}
It is the clutter type (or land cover type) of the \ac{UE} (For example: open street, dense buildings, sparse buildings, trees, water etc.). Each clutter type has its own effect on the signal and this feature tries to learn this behavior. It is denoted by $c_{\textit{UE}}$.

\subsubsection{Number of Building Penetrations in each Clutter Type}
This is the number of buildings penetrated by the signal in each unique clutter in the direct path between a BS and UE. Different clutters can be different types of buildings, each having different penetration loss (dB). If our observation area consists of 15 different clutter classes, then this feature is subdivided into 15 different features, each representing the number of building penetrations in that respective clutter, whose sum equals the total number of building penetrations in the propagation path of that UE. It is denoted by $N_{c}$.

\subsubsection{Indoor Distance in each Clutter Type}
Indoor distance (in meters) covered by each unique clutter in the direct path between a BS and UE. This feature characterizes the linear loss (dBm/m) experienced by the signal in different indoor environments. Again, this feature is subdivided into the total number of clutters in the observation area, whose sum equals the total indoor distance in the propagation path of that UE. It is denoted by $d_{\textit{indoor}_{c}}$.

\subsubsection{Outdoor Distance in each Clutter Type}
Outdoor distance (in meters) covered by each unique clutter in the direct path between a BS and UE. This feature characterizes the linear loss (dBm/m) experienced by the signal in different outdoor environments. Again, this feature is subdivided into the total number of clutters in the observation area, whose sum equals the total outdoor distance in the propagation path of that UE. It is denoted by $d_{\textit{outdoor}_{c}}$.

\begin{table*}[!t]
        \begin{center}
                \caption{Key Symbol Definitions}
                \label{tab:symbol}
                \begin{tabular}{|c|c|c||c|c|c|}
                        \hline
                        \thead{Symbol} & \thead{Units} & \thead{Definition} & \thead{Symbol} & \thead{Units} & \thead{Definition}\\
                        \hline
                        $\theta_{\textit{BS}}$ & $^o$ & Azimuth Angle of (BS) Departure & 
                        $P_{\textit{BS}}$ & dBm & BS Transmit Power\\
                        \hline
                        $\phi_{\textit{BS}}$ & $^o$ & Tilt Angle of (BS) Departure & 
                        $P_{\textit{UE}}$ & dBm & RSS of a UE\\
                        \hline
                        $\theta_{\textit{UE}}$ & $^o$ & Azimuth Angle of (UE) Arrival & 
                        $d$ & $m$ & Propagation Distance \\
                        \hline
                        $\phi_{\textit{UE}}$ & $^o$ & Tilt Angle of (UE) Arrival & 
                        $d_{\textit{vert}}$ & $m$ & 
                        Effective BS Height\\
                        \hline
                        $\phi_{\textit{ver}}$ & $^o$ & Vertical Angular Separation & $d_{\textit{man}}$ & $m$ & Manhattan Distance\\
                        \hline
                        $\theta_{\textit{hor}}$ & $^o$ & Horizontal Angular Separation &
                        $L$ & - & LoS/NLoS State \\
                        \hline
                        $f$ & MHz & Operating Frequency & 
                        $d_{\textit{FD}}$ & $m$ & Distance from BS to First Diffraction Point\\
                        \hline
                        $h_{\textit{BS}}$ & $m$ & BS Antenna Height & $d_{\textit{LD}}$ & $m$ & Distance from BS to Last Diffraction Point\\
                        \hline
                        $h_{\textit{UE}}$ & $m$ & UE Antenna Height & 
                        $N$ & - & Number of Building Penetrations\\
                        \hline
                        $c_{\textit{BS}}$ & int & BS Clutter Type &
                        $d_{\textit{indoor}}$ & $m$ & Indoor Distance in the Propagation Path\\
                        \hline
                        $c_{\textit{UE}}$ & int & UE Clutter Type &  
                        $d_{\textit{outdoor}}$ & $m$ & Outdoor Distance in the Propagation Path\\
                        \hline
                        $D_{\textit{DHM}}$ & - & Raster Grid Data of Digital Height Model &
                        $N_{c}$ & - & No. of Building Penetrations in each Unique Clutter\\
                        \hline
                        $D_{\textit{DTM}}$ & - & Raster Grid Data of Digital Terrain Model & 
                        $d_{\textit{indoor}_{c}}$ & $m$ & Indoor Distance in each Unique Clutter\\
                        \hline
                        $D_{\textit{DLU}}$ & - & Raster Grid Data of Digital Land Use Map &
                        $d_{\textit{indoor}_{c}}$ & $m$ & Outdoor Distance in each Unique Clutter\\
                        \hline
                \end{tabular}
        \end{center}
\end{table*}
\vspace{10pt}
\subsection{RSS Modeling using Machine Learning Methods}
RSS prediction is essentially a regression problem, where the key features proposed earlier are used as input for training ML models, to learn the complex behavior of a signal passing through a wireless channel. \cref{alg:rss} explains the process of removing randomness from the UE traces (to some extent) by gridding (averaging all measurements in a spatial bin) and then training the ML model using the computed key features as input and the corresponding expected value of RSS as output.

In this paper, we investigate a range of different \textit{Parametric}, \textit{Non-Parametric} and Ensembles of machine learning regression algorithms for their strengths and weaknesses while modeling RSS and implementing them.
\begin{itemize}
    \item \uline{Parametric algorithms}, such as Linear Regression, assumes training data to be of a specific functional form with a fixed size of parameters.
    \item \uline{Non-Parametric algorithms}, on the other hand, such as k-Nearest Neighbors, Decision Tree and Neural Networks, are free to assume any functional form of the training data.
    \item \uline{Ensemble algorithms} are of two types: Bagging and Boosting, which further have several variants.
\end{itemize} 
Earlier works on propagation modeling \cite{dai2018propagation} were mostly based on parametric models and some ensemble learning models.

\begin{algorithm}
        \caption{RSS Prediction Model Training Algorithm}\label{alg:rss}
        \begin{algorithmic}[1]
                \Input $D_{\textit{UE}}$, $D_{\textit{BS}}$, $D_{\textit{DTM}}$, $D_{\textit{DHM}}$, $D_{\textit{DLU}}$
                \Output RSS Prediction Model $\mathbfcal{M}(F)$
                \For{all UE traces}
                \State \algparbox{Map its location to pre-defined grids (e.g., 10m x 10m)}
                \EndFor
                \For{each unique grid}
                \For{each unique serving BS}
                \State \algparbox{\algparbox{Average out the RSS ($P_{\textit{UE}}$) of all users to offset randomness}}
                \State \algparbox{\algparbox{Compute feature vector $F=[d, \theta_{\textit{hor}},
                \phi_{\textit{ver}},
                d_{\textit{vert}},\\
                d_{\textit{man}}, 
                L, d_{\textit{FD}}, d_{\textit{LD}}, N, d_{\textit{indoor}}, d_{\textit{outdoor}}, c_{\textit{BS}},
                c_{\textit{UE}}, N_{c},\\ d_{\textit{indoor}_{c}},  d_{\textit{outdoor}_{c}}]$}}
                \EndFor
                \EndFor
                \State Train the Machine Learning model $\mathbfcal{M}$ using Feature Matrix $\mathbfcal{F}$, whose each row corresponds to a feature vector $F$
                \State \Return $\mathbfcal{M}(F)$
        \end{algorithmic}
\end{algorithm}

\subsubsection{Criteria for Model Evaluation and Selection}
In this work, we have done a comprehensive and multi-faceted performance evaluation of the state-of-the art ML algorithms, that includes \textit{predictive performance}, \textit{generalization performance} and \textit{computational performance}, and also provided insights from each of these algorithms to make this paper self-contained.

\begin{enumerate}

\item \uline{Predictive Performance:} The predictive performance of a model indicates the model accuracy for unseen data. In our analysis, we have used \acf{RMSE} and \acf{$R^2$} performance metrics defined below to judge the predictive performance of models.
\begin{equation*}
RMSE = \sqrt{\frac{\Sigma_{i=1}^{N}{(P_{\textit{UE}_i} -\hat{P}_{\textit{UE}_i})^2}}{N}},
\end{equation*}

\begin{equation*}
R^2 = 1-\frac{{SS}_{res}}{{SS}_{total}}=1-\frac{\Sigma_{i}^{}{(P_{\textit{UE}_i} -\hat{P}_{\textit{UE}_i})^2}}{\Sigma_{i}^{}{(P_{\textit{UE}_i} -\bar{P}_{\textit{UE}_i})^2}}.
\end{equation*}

\vspace{3pt}
Here $P_{\textit{UE}}$ is the actual \ac{RSS} of a UE, $\hat{P}_{\textit{UE}}$ is the predicted \ac{RSS} of a UE, $\bar{P}_{\textit{UE}}$ is the mean value of \ac{RSS} and $N$ is the number of UE traces in the test data. ${SS}_{res}$ and ${SS}_{tot}$ corresponds to the \textit{residual sum of squares} and \textit{total sum of squares}, respectively. $RMSE$ is measured here in dB, whereas $R^2=1$ in the best case and $R^2=0$ when the model output is always equal to its mean value in test data. In rare scenarios, $R^2<0$, when the model predictions are even worse than the baseline mean value prediction.
\vspace{5pt}
\item \uline{Generalization Performance:} The generalization performance of a model indicates its robustness in predictive performance for unseen data (or scenarios). In our analysis, we have used 5-fold repeated cross-validation technique along with its mean and standard deviation value for all iterations to judge the generalization ability of a model on unseen data (propagation scenarios) with a certain confidence, even when using very sparse training data (2\% in our analysis). In other words, the generalization ability of the model can be judged both by the standard deviation of its mean RMSE or the increase in its predictive performance when using sparse training data for all cross-validation iterations.
\vspace{5pt}
\item \uline{Computational Performance:} Computational performance of a ML model can be judged by its training time, which indicates the time and therefore resources it takes to train the model and prediction time, which shows its prediction latency. These values are extremely crucial in a production setting where we have cost (or resources) and latency constraints. In our analysis, all the ML methods are evaluated using the same number of CPU cores for a fair comparison.
\end{enumerate}

\subsubsection{Model Evaluation}

\begin{enumerate}

\item \uline{Linear Regression:}
As evidenced by our experiments, linear regression method \cite{altman2015simple} doesn't seem to be suitable to capture the complex non-linear nature of the wireless channel. In our results, we have shown (in \cref{fig:ml_compare}) that it gives a high prediction \ac{RMSE} of \textit{5.45 dB} and a low \ac{$R^2$} score of \textit{0.65}.

\vspace{3pt}
\item \uline{k-Nearest Neighbors:}
\ac{k-NN} \cite{altman1992introduction} doesn't seem to handle non-linearity well in case of sparse training data, as the prediction in test data is basically the mean of $k$ nearest data points in the training data. Also, it has the highest computation cost at run-time among the tested algorithms, as evidenced in results (\cref{fig:ml_compare}).

\vspace{3pt}
\item \uline{Decision Tree:}
A single \ac{DT} \cite{breiman1984classification} in our experiments, is unable to generalize well, especially with sparse training data and seems to overfit, therefore also doesn't seem to be a suitable choice for a \ac{ML}-based propagation model.

\vspace{3pt}
\item \uline{Random Forest:}
Random Forest \cite{breiman2001random} is an Ensemble learning method, which combines several decision trees, using \textit{Bootstrap Aggregation (or Bagging)} technique, to improve the predictive performance of a single decision tree. Here, each tree is trained on a random subset, with replacement, of training data. Also, each node is split among a random subset of input features, which may not be the best split among all features. This randomness increases the bias of the forest, when compared to a single non-random tree. The output here is the average prediction of all individual trees, and due to this averaging, variance in the ensemble model decreases, which more than make up for the increase in bias, hence improving performance of the overall model, \textit{RMSE of 3.46 dB} as compared to 4.76 dB in a single decision tree. Another advantage is that, as opposed to a single decision tree, random forest is robust to outliers in the training data. The main drawback of using this method is the slow prediction speed, as evidenced in our results (See \cref{fig:ml_compare}), due to a large number of trees, making it unsuitable for real-time predictions.

\vspace{3pt}
\item \uline{Extremely Randomized Trees:}
Extremely Randomized Trees is another \textit{Bagging Ensemble} learning method, which goes one step further in randomizing the trees, as compared to Random Forest. In addition to each tree trained on a random subset of data and best split at each node chosen on a random subset of features, thresholds are also picked at random for every candidate feature at a node. This increase in randomness, further decreases the variance of the ensemble model, at the cost of slight increase in bias. It has all the pros of Random Forest, plus a reduction in training time, but the major drawback is still high prediction time, as evidenced in our results (See \cref{fig:ml_compare}).

\vspace{3pt}
\item \uline{Adaptive Boosting:}
Adaptive Boosting (AdaBoost) is a  \textit{Boosting Ensemble} model. In boosting, models are built in sequence, so that each subsequent model learns from the mistakes of the previous model, to create a stronger model. In AdaBoost, each subsequent model is forced to focus on samples which were badly predicted in the previous model. This is done by giving higher weights to those samples in the training set, whose prediction error was high in the previous model, and vice versa. Weighted sampling is then used in the subsequent model to generate a derived training set, using sampling with replacement. The probability that a training example appears in the training set is relative to its weight. The final output is a weighted average of all the model's output, where more weight is placed on stronger models. Consequently, \textit{the bias of the combined model is reduced in boosting, as opposed to bagging, where the variance was reduced,} by averaging several weak learners. As shown in \cref{fig:ml_compare}, bagging methods outperforms AdaBoost in terms of higher prediction accuracy. The other disadvantage of this technique is that it cannot be parallelized, since the training of each subsequent tree model, depends on the output of previous model, therefore, its training time is much higher as compared to bagging ensemble methods.

\vspace{3pt}
\item \uline{Gradient Boosting Decision Trees:}
Gradient Boosting Decision Trees (GBDT) is another Boosting ensemble model, which works on the same Boosting principle of learning from the previous model's mistake, but the difference lies in how to learn from the mistakes of previous model. \textit{Gradient Boosting learns from the error residuals (gradients) of the previous model directly}, unlike AdaBoost, which changes the sample distribution at every iteration, by giving higher weights to under-fitted (or badly predicted) samples. The goal is to iteratively minimize the prediction error, by training each subsequent decision tree model, on the residual errors (prediction errors) made by the previous model, this process is essentially a gradient descent optimization on the overall composite model. The final output would then be the sum of predictions from all the models. Our results show that Gradient Tree Boosting has much better prediction performance (\textit{RMSE of 4.32 dB}) as compared to AdaBoost (See \cref{fig:ml_compare}).

\vspace{3pt}
\item \uline{Extreme Gradient Boosting:}
Extreme Gradient Boosting (XGBoost) \cite{chen2016xgboost} is an advanced implementation of gradient boosting and follows the same principle. The main advantage of XGBoost is the ability of parallel processing, therefore much faster as compared to GBDT. While it's not possible to create trees in parallel because each tree is dependent on the previous, it's possible to build a tree using all the cores, by building several nodes within each depth of a tree, in parallel. To improve the performance of the model, \textit{Weighted Quantile Stretch} idea is used to reduce the search space while finding the best split, by looking at the distribution of features across all instances in a leaf. It further reduces the computational complexity by learning the sparsity patterns in the data and skip samples with missing values while making a split. Moreover, it includes \textit{regularization} to prevent over-fitting and improve overall performance of the model (See \cref{fig:ml_compare}), due to which it is also sometimes called as \textit{regularized gradient boosting}.

\vspace{3pt}
\item \uline{Light Gradient Boosting:}
Light Gradient Boosting Machine (LightGBM) \cite{ke2017lightgbm} is another implementation of Gradient Boosting, which improves on XGBoost. It can train on larger datasets in a fraction of time and with comparable accuracy, as compared to XGBoost, hence the word \textit{Light} is used. It uses a technique called \textit{Gradient-based One-Side Sampling (GOSS)} to intelligently extract the most useful information as fast as possible, by randomly skipping the samples with less information (small gradients) in the dataset. Moreover it has introduced another method called \textit{Exclusive Feature Bundling (EFB)} for reducing model complexity, by combining similar features in a near lossless way. Our results have shown that LightGBM has better accuracy as compared to XGBoost in sparse training data scenario and 12x faster training speed (See \cref{fig:ml_compare}).

\vspace{3pt}
\item \uline{Categorical Boosting:}
Categorical Boosting (CatBoost) \cite{prokhorenkova2018catboost} is a gradient boosting algorithm whose power lies in processing categorical features in the dataset. Categorical features have values which are discrete and not related to each other. CatBoost incorporates several innovative methods to deal with these features at training, instead of during data pre-processing. Furthermore, it incorporates a modified gradient boosting algorithm called \textit{ordered boosting}, to avoid \textit{target leakage} present in standard gradient boosting algorithms, as they rely on the target of all training samples at each iteration, resulting in biasness. Here, however, training is done on independent random permutations of the dataset at each iteration, to avoid this \textit{prediction shift}. Therefore, CatBoost can outperform other gradient boosting algorithms, specially if you have categorical variables in the data (for instance, LOS State, BS and UE Clutter Types are the categorical features used in our model). As shown in our results (\cref{fig:ml_compare}), prediction RMSE is reduced to \textit{3.74 dB} at the cost of increase in training time. It is worth mentioning here that the GPU implementation of this algorithm is faster than both XGBoost and LightGBM, but in our results we have used CPU for training these algorithms.

\vspace{3pt}
\item \uline{Deep Neural Network:}
Deep Neural Network (DNN) algorithm belongs to a special class of machine learning, called \textit{deep learning} and creates a \textit{multi-layer perceptron (MLP)} to find the input-output associations. Its basic structure consists of an input layer, output layer and one or more hidden layers between them, each containing several neurons (or nodes). Neurons in the input layer equals the number of input features, whereas output layer consists of one neuron which holds the prediction output. Number of hidden layers and its neurons are variable, and depends on the complexity of model it is trying to learn. Our extensive investigations on DNN design show that, for learning the behavior of RSS in a wireless channel, 6 hidden layers each consisting of 32 neurons provide the most optimal results, any increase or decrease in this number results in over-fitting or under-fitting on the training data, respectively. The DNN used in this study has \textit{Rectified Linear Unit (ReLu)} activation function in the hidden layers, whereas output layer uses \textit{linear} activation function. In our simulation results (\cref{fig:ml_compare}), DNN performs worse than ensemble-based methods and also has the highest computational cost.
\end{enumerate}

\subsubsection{Model Selection}
For selecting the best performing model, we should overall look at the predictive, generalization and computational performance across all evaluated models. For a fair comparison, all the ML methods are evaluated using the same number of CPU cores. Furthermore, in all experiments, 5-fold repeated cross validation is used so that the results are generalizable in all propagation scenarios.

\begin{figure}
        \centering     
        \subfigure[Training time (normalized by the highest value of DNN) and Prediction Time (normalized by the highest value of k-NN) ]{\includegraphics[width=.98\linewidth]{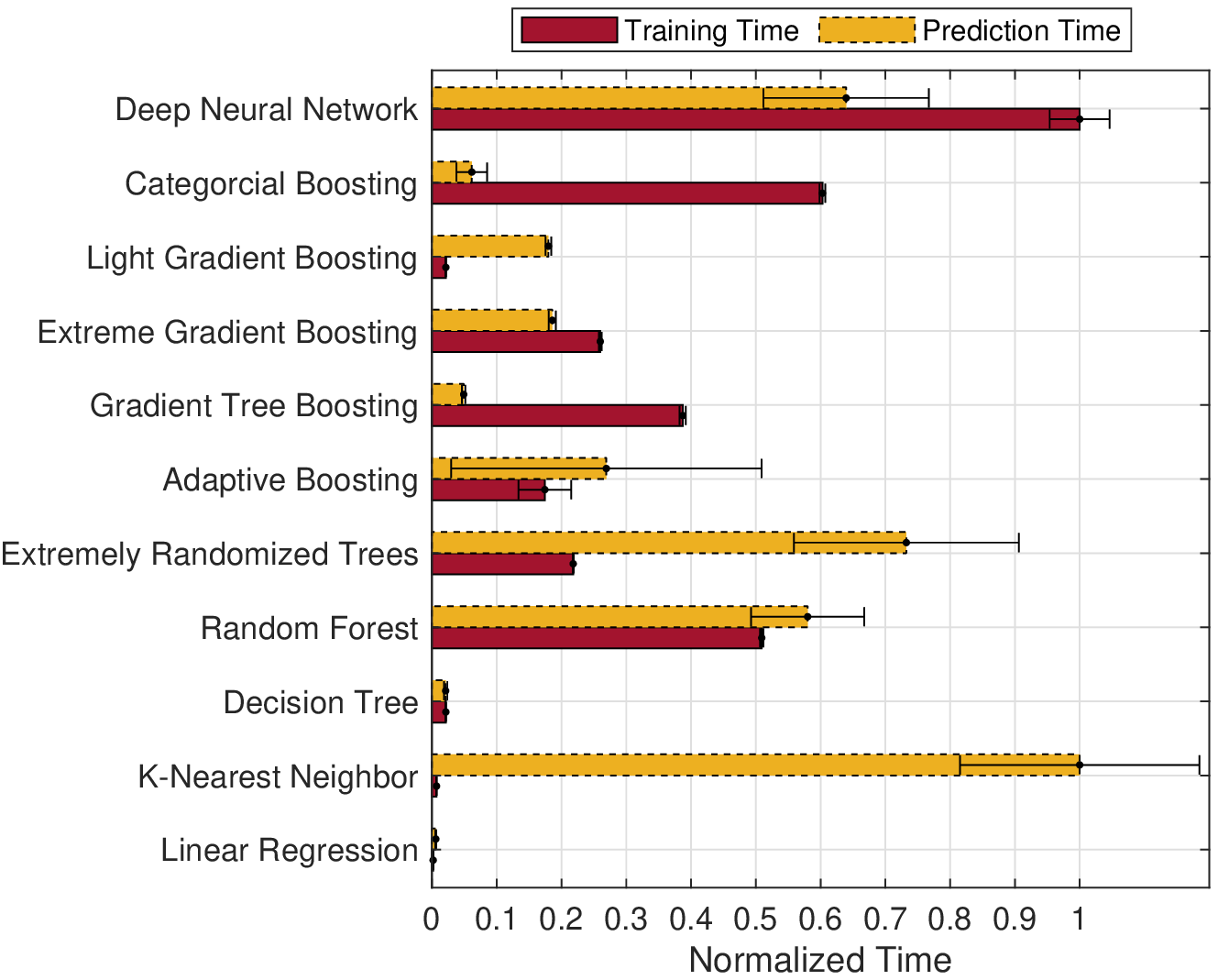}\label{fig:ml_compare_a}}
        \subfigure[RSS Prediction Error]{\includegraphics[width=.98\linewidth]{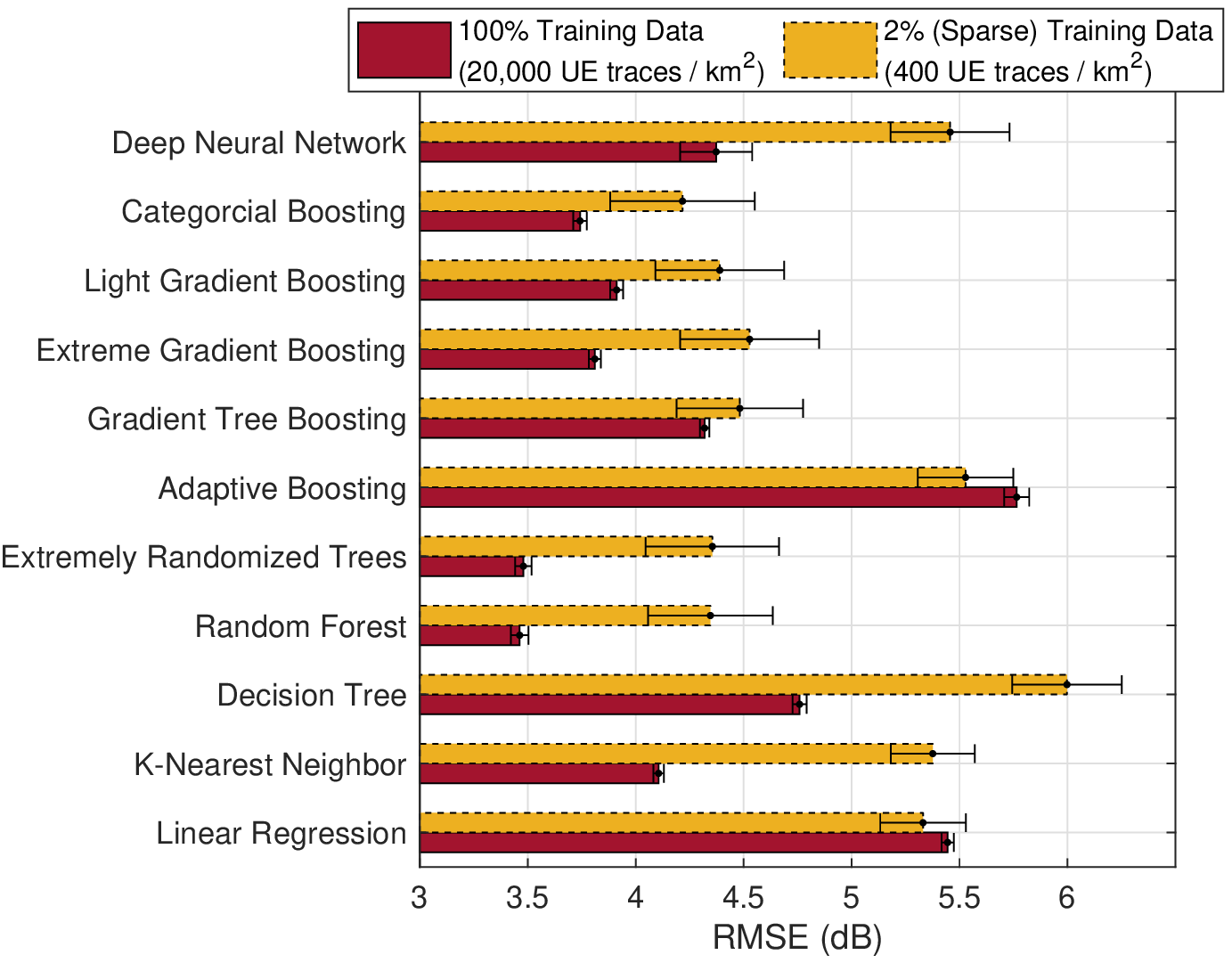}\label{fig:ml_compare_b}}
        \subfigure[Model R-Squared Value]{\includegraphics[width=.98\linewidth]{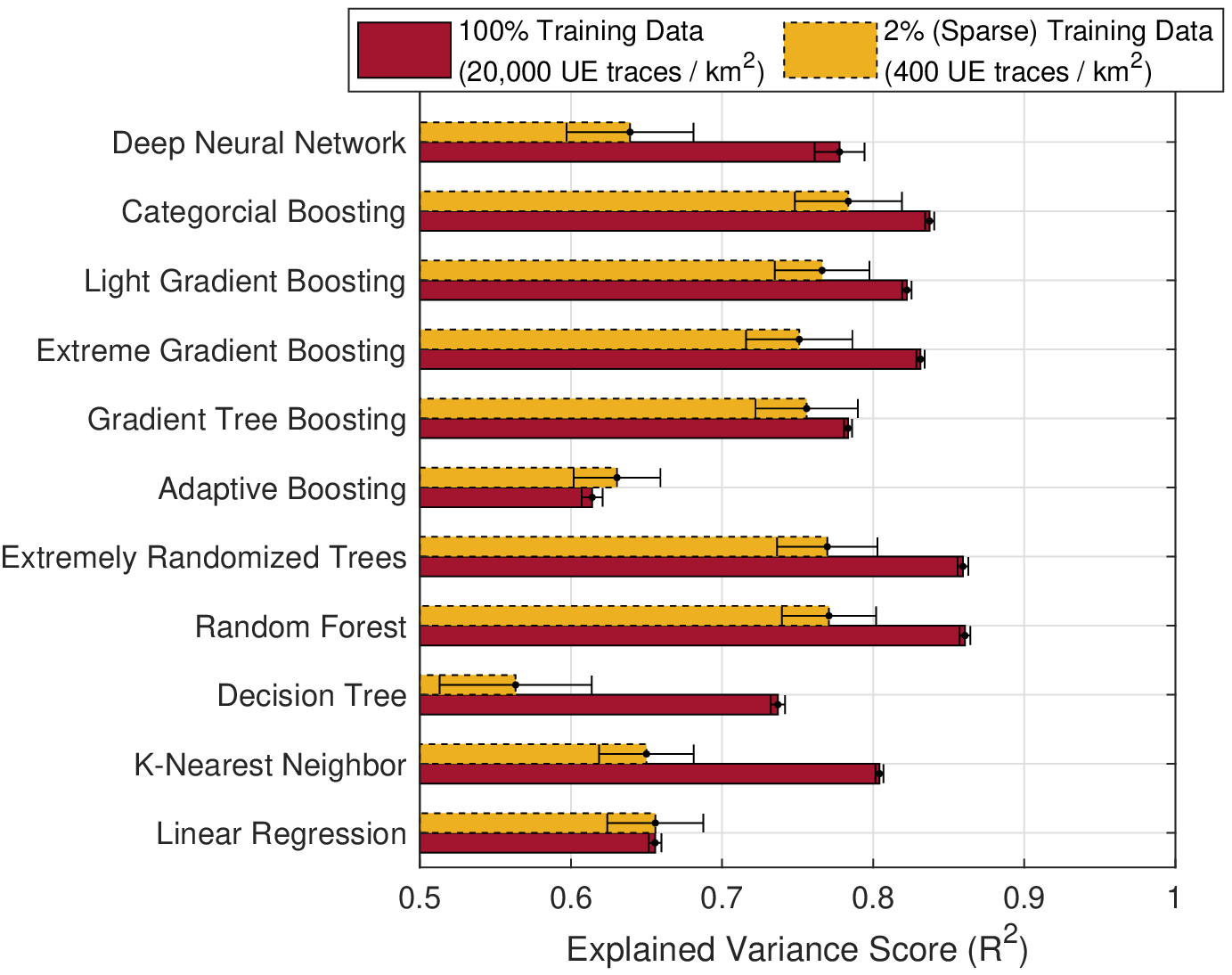}\label{fig:ml_compare_c}}
        \caption{Comparison of various Machine Learning Algorithms w.r.t ({\bf a}) Training Time, Prediction Time, ({\bf b}) Prediction Error, ({\bf c}) R-Squared Value and Robustness to Sparsity of Training Data, for Modeling RSS (Height of bars represent the mean value and Error bar represent the standard deviation using 5-fold Repeated Cross Validation)}
        \label{fig:ml_compare}
\end{figure}

In \cref{fig:ml_compare_a}, training and prediction time of all the methods are normalized to the highest value individually. DNN has the highest, whereas linear regression has the lowest training and prediction time. In \cref{fig:ml_compare_c}, comparison of \ac{$R^2$} value is given, where CatBoost and LightGBM algorithms perform the best in capturing the variance of RSS (or pathloss) and learning complex non-linear relationships in a wireless channel and have the lowest Root Mean Squared Error (RMSE) in sparse training data scenarios (shown in \cref{fig:ml_compare_b}), whereas linear regression has the highest RMSE for RSS prediction as the complex non-linear nature of wireless channel renders it unsuitable.

All the models are also separately trained on only 2\% of training data, to evaluate their performance in case of data sparsity, as is the case in real practical scenarios. DNN shows the highest impact (loss in accuracy) due to data sparsity.

\textit{Overall, LightGBM algorithm outperforms others, especially for real-time implementation, due to its lightning fast training process}. RSS model trained using LightGBM algorithm is used for further simulations and results.

\subsubsection{Model Improvement using Hyperparameter Optimization}
The baseline LightGBM model performance shown in \cref{fig:ml_compare} is further improved by optimizing its hyperparameters for the RSS (or pathloss) prediction task, according to the special characteristics of the wireless channel. 

\begin{figure}
        \centering     
        \subfigure[Grid Search]{\includegraphics[width=.49\linewidth]{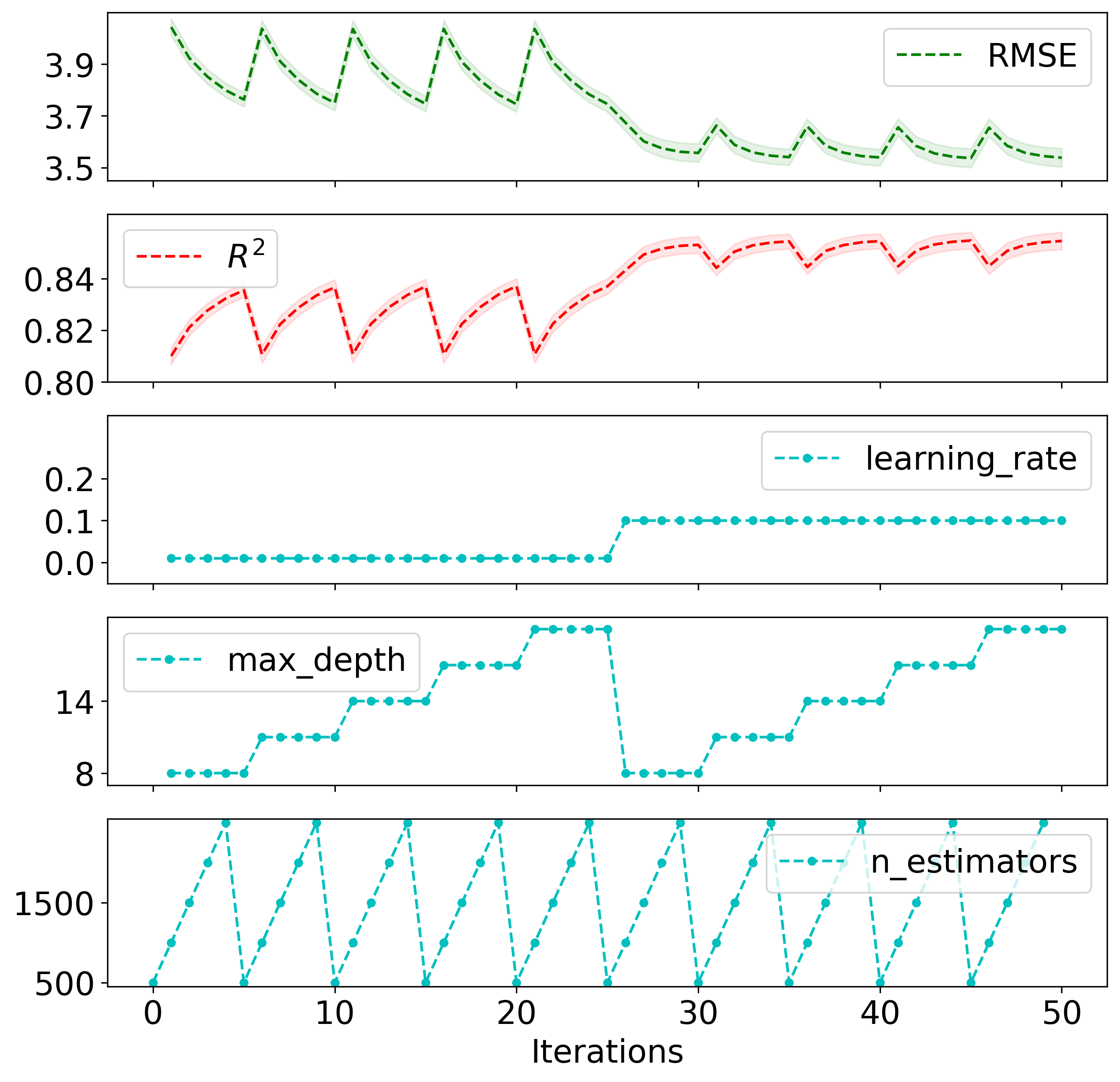}\label{fig:gs}}
        \subfigure[Random Search]{\includegraphics[width=.49\linewidth]{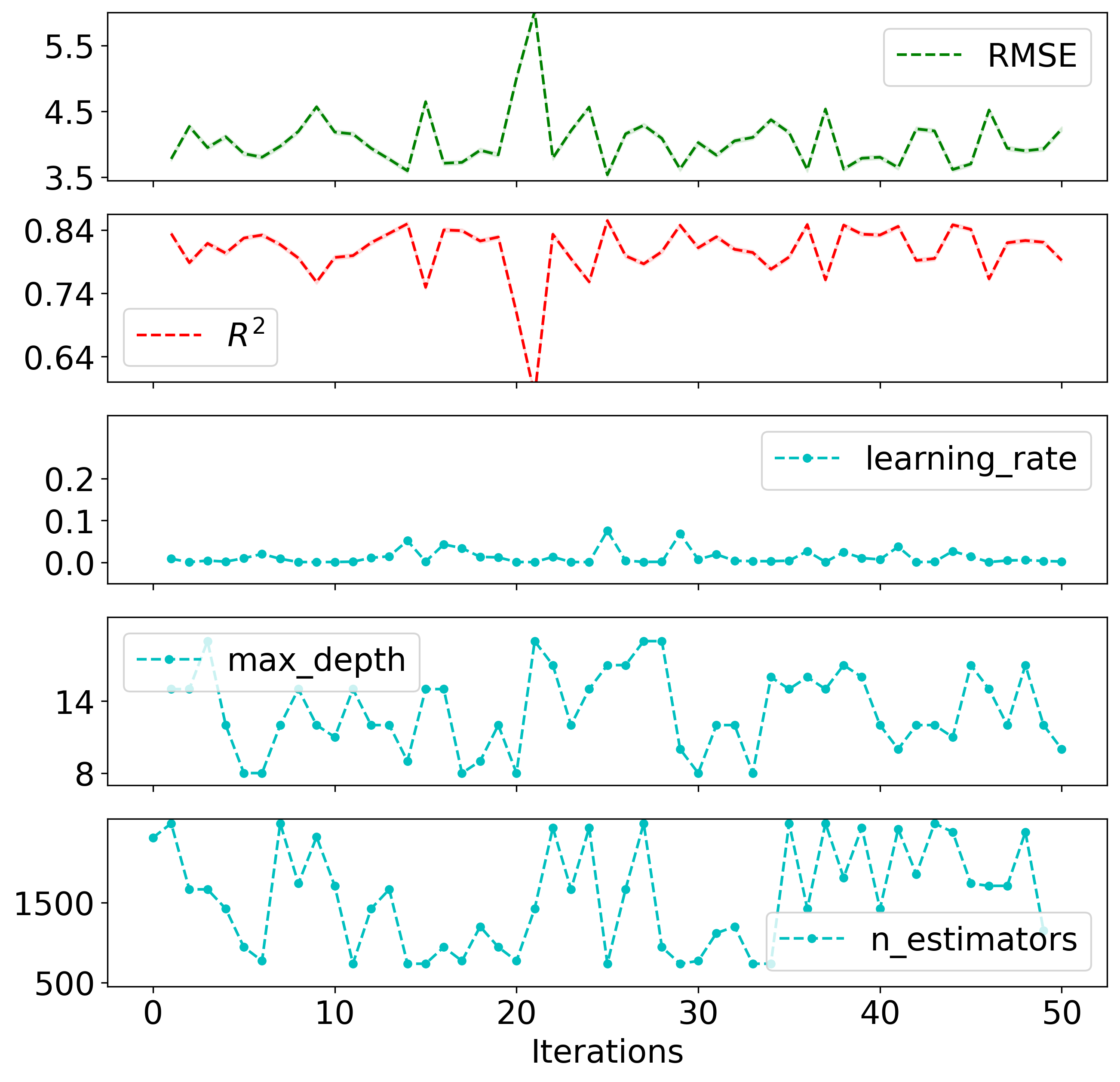}\label{fig:rs}}
        \newline
        \subfigure[Bayesian TPE]{\includegraphics[width=.49\linewidth]{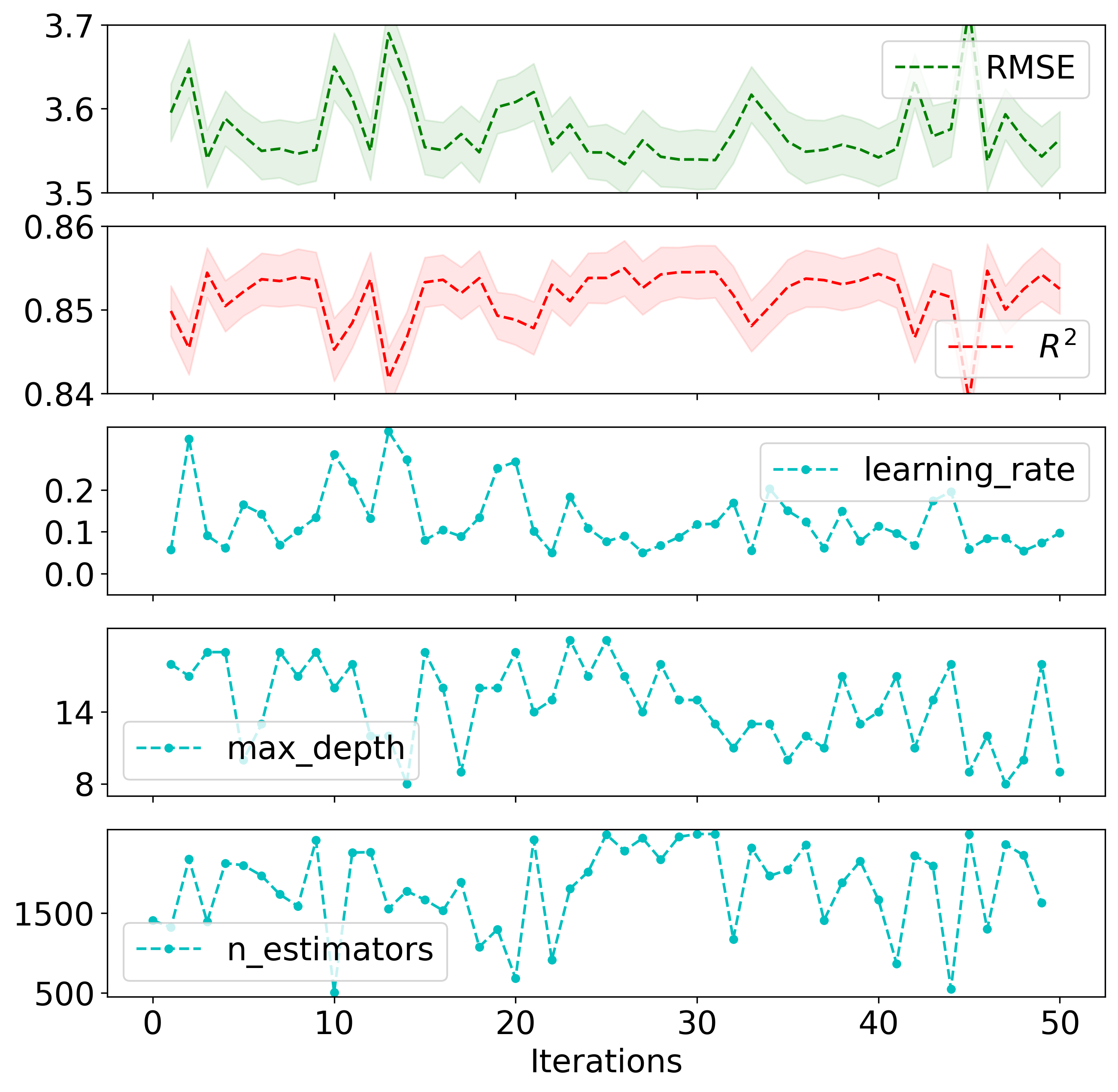}\label{fig:tpe}}
        \subfigure[Simulated Annealing]{\includegraphics[width=.49\linewidth]{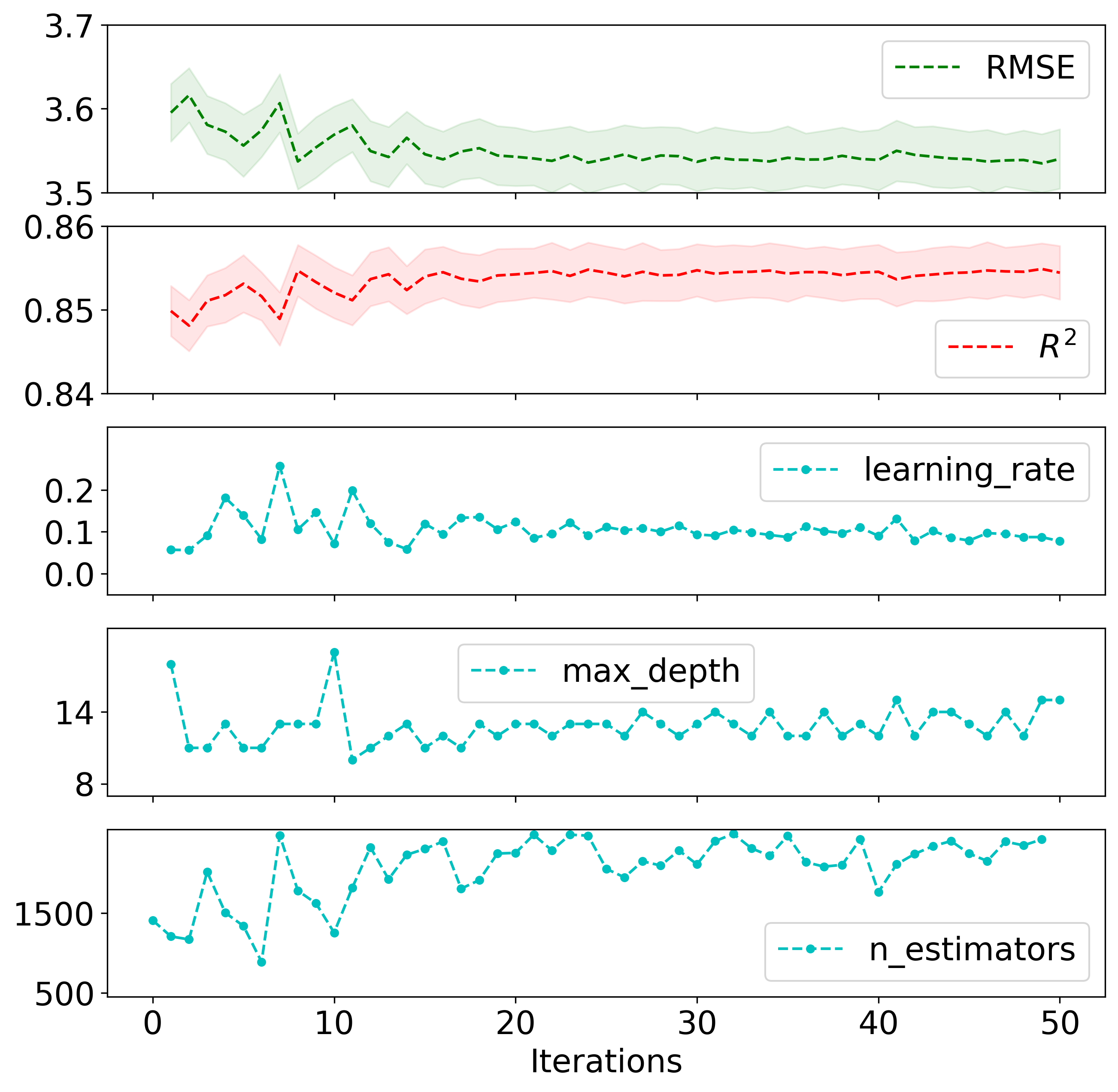}\label{fig:sa}}
        \subfigure[Comparison of hyperparameter tuning approaches w.r.t. Performance Gain and Convergence Time. Bayesian optimization performed the best here by achieving a 10\% improvement in prediction RMSE as compared to baseline LightGBM model, in just 3 search iterations]{\includegraphics[width=.79\linewidth]{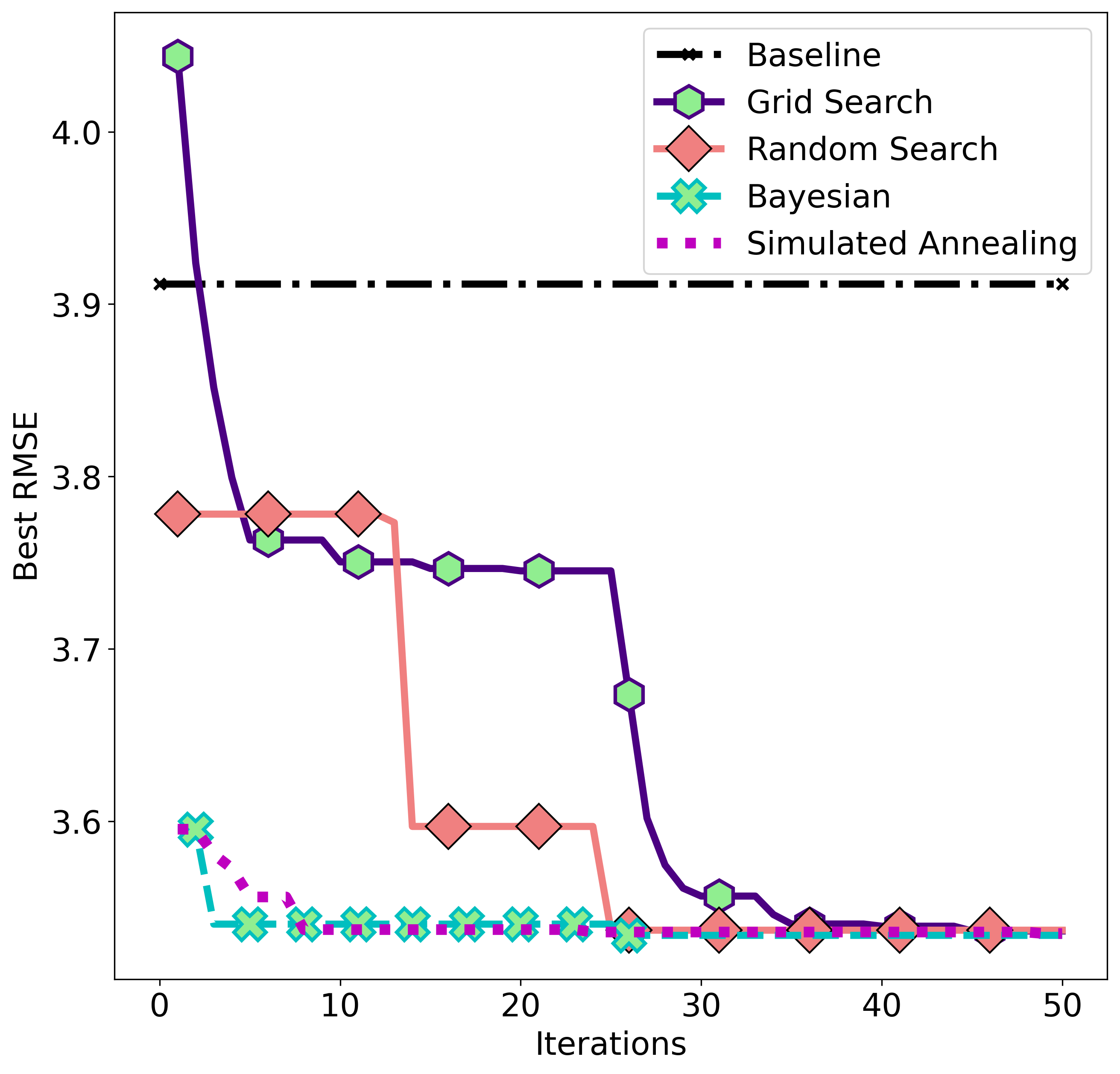}\label{fig:tuning_compare}}
        
        \caption{Comparison of different Hyper-parameter tuning approaches for improving the performance of baseline LightGBM model, in terms of RSS Prediction RMSE, $R^2$ and convergence time (In subplots (a)-(d), RMSE and $R^2$ of the LightGBM model is plotted against different combinations of hyperparameters, after performing 5-fold Repeated Cross Validation at each search iteration of the tuning process (dotted line/curve represents the Mean value and filled area/polygon around it represents the standard deviation using 5-fold Repeated Cross Validation)}
        \label{fig:model_tuning}
\end{figure}

The hyperparameters selected to be optimized are: 1) `number of estimators' in the range of $500-2500$, 2) `maximum tree depth' in the range of $5-20$ and 3) `learning rate' in the range of $0.1-0.001$. Furthermore, 5-fold repeated cross validation is used for each combination of hyperparameters and model's performance is evaluated based on its \ac{RMSE} and \ac{$R^2$}. Four different hyperparameter optimization approaches are evaluated (shown in \cref{fig:model_tuning}) in terms of performance gain and convergence time:

\begin{enumerate}
\item \uline{Grid Search:}
This approach does an exhaustive search over the entire search space of hyperparameters. \cref{fig:gs} shows the mean \ac{RMSE} and \ac{$R^2$} of LightGBM model at different combinations of hyperparameters. The model converges to its best RMSE after 50 iterations (as shown in \cref{fig:tuning_compare}).

\vspace{5pt}
\item \uline{Random Search:}
This approach also does an exhaustive search over the entire search space of hyperparameters, but picks them randomly, therefore its convergence time is more likely to be less than grid search method, as shown in \cref{fig:tuning_compare}, where the model converges to its best RMSE after 25 iterations.

\vspace{5pt}
\item \uline{Bayesian:}
In this approach, hyperparameters are tuned using a Bayesian optimization algorithm, known as Tree-structured Parzen Estimator (TPE) \cite{bergstra2011algorithms}. This Bayesian approach is a model-based approach, and as search iterations progresses, it switches from exploration to exploitation to minimize the objective function loss (concentrating on the hyperparameter combinations that resulted in lower loss, which in our case is the RMSE). This approach sometimes gets trapped in the local minima of the objective function, an issue which is not faced by grid or random search. \cref{fig:tpe} and \cref{fig:tuning_compare} shows the superior performance of this approach as it converges in only 3 search iterations.

\vspace{5pt}
\item \uline{Simulated Annealing:}
This approach is a meta-heuristic optimization algorithm \cite{van1987simulated}, that is simpler and is preferred over its Bayesian counterpart when the objective function is simple to evaluate. But it seems to converge slowly than the Bayesian approach (as evidenced in \cref{fig:tuning_compare}, where it took 8 iterations to converge). After these 8 iterations, our proposed ML algorithm shows a significant performance gain, as its prediction RMSE is reduced to 3.54 dB, as compared to 3.91 dB earlier in the baseline LightGBM algorithm using the default hyperparameters.

\end{enumerate}

\section{Comparison with Empirical Radio Propagation Models}\label{section3}
We also compare the performance of our proposed AI-driven 3D propagation model based on improved LightGBM algorithm with traditional empirical propagation models, as they are currently used in state-of-the-art commercial planning tools to characterize the propagation behavior of a radio signal in different conditions. Empirical models offer a mathematical equation to calculate the path loss at any given point from the BS, and are based on data collected in a specific scenario.

\subsection{COST-Hata Model}
It is an empirical model for pathloss calculation \cite{1991urban}, that extends the Hata formulae \cite{hata1980empirical} to frequencies upto 2 GHz and it also takes into account the topo map (DTM) between the BS and UE and morpho map (DLU) only at the receiver. The below equation is valid for urban environments with 1.5 m UE height.
\begin{eqnarray}
\nonumber
\label{eq:cost_hata1}
L_{\textit{path}}=A_1+A_2*log(f)+A_3*log(h_{\textit{BS}})+\\(B_1+B_2*log(h_{\textit{BS}})+B_3*h_{\textit{BS}})*log(d).
\end{eqnarray}

Here $L_{\textit{path}}$ is the pathloss (in dB), $A_1=46.3$, $A_2=33.9$, $A_3=-13.82$, $B_1=44.9$, $B_2=-6.55$, $B_3=0$ are user-defined parameters, $f$ is the carrier frequency (in MHz), $h_{BS}$ is the height of BS and $d$ is the propagation distance between BS and UE.

For Urban Areas:
\begin{eqnarray}
\nonumber
\label{eq:cost_hata2}
L_{\textit{path}}^\prime=L_{\textit{path}}-a(h_{\textit{UE}}).
\end{eqnarray}

For Sub-Urban Areas:
\begin{eqnarray}
\nonumber
\label{eq:cost_hata3}
L_{\textit{path}}^\prime=L_{\textit{path}}-a(h_{\textit{UE}})-2*(log(\frac{f}{28}))^2-5.4.
\end{eqnarray}

For Quasi-Open Rural Areas:
\begin{eqnarray}
\nonumber
\label{eq:cost_hata4}
L_{\textit{path}}^\prime=L_{\textit{path}}-a(h_{\textit{UE}})-4.78*(log(f))^2+18.33*log(f)\\-35.94.
\nonumber
\end{eqnarray}

For Open Rural Areas:
\begin{eqnarray}
    \nonumber
    \label{eq:cost_hata5}
    L_{\textit{path}}^\prime=L_{\textit{path}}-a(h_{\textit{UE}})-4.78*(log(f))^2+18.33*log(f)\\-40.94.
    \nonumber
\end{eqnarray}

Where $L_{\textit{path}}^\prime$ is the corrected pathloss and $a(h_{\textit{UE}})$ is the correction factor for UE height different from 1.5 m.
\\

For Rural/Small Cities:
\begin{eqnarray}
\nonumber
\label{eq:cost_hata6}
a(h_{\textit{UE}})=(1.1*log(f)-0.7)*h_{\textit{UE}}-(1.56*log(f)-0.8).
\nonumber
\end{eqnarray}

For Open Rural Areas:
\begin{eqnarray}
\nonumber
\label{eq:cost_hata7}
a(h_{\textit{UE}})=3.2*(log(11.75*h_{\textit{UE}}))^2-4.97.
\nonumber
\end{eqnarray}

\subsection{Stanford University Interim (SUI) Model}
It is derived from the Erceg-Greenstein propagation model \cite{erceg1999empirically} and is valid for 1900-6000 MHz. It also takes into account the topo map (DTM). It uses the following formula:
\begin{eqnarray}
\label{eq:sui1}
\nonumber
L_{\textit{path}}=-7366+26*log(f)+10*a(h_{\textit{BS}})*(1+log(d))\\
-a(h_{\textit{UE}}),
\end{eqnarray}
where,
\begin{eqnarray}
\label{eq:sui2}
\nonumber
a(h_{\textit{BS}})=a-b*h_{\textit{BS}}+\frac{c}{h_{\textit{BS}}},\\
\nonumber
a(h_{\textit{UE}})=X*log\left(\frac{h_{\textit{UE}}}{2}\right).
\end{eqnarray}
\\
Here $a(h_{\textit{BS}})$ and $a(h_{\textit{UE}})$ are the correction factors for BS and UE antenna heights, respectively, $f$ is the operating frequency and $d$ is the propagation distance (in km). $a=4.6$, $b=0.0075$, $c=12.6$ and $X=10.8$ are the correction constants which depend on the terrain type \cite{erceg2001channel}.

\subsection{Standard Propagation Model (SPM)}

It is derived from the Hata formulae and is valid for 150-3500 MHz. It also takes into account the topo map (DTM) and morpho map (DLU) between the BS and UE. It is given by the following formula:
\begin{eqnarray}
\label{eq:spm1}
\nonumber
L_{\textit{path}}=K_1+K_2*log(d)+K_3*log(h_{\textit{BS}}^{\prime})+K_4*L_{\textit{diff}}\\
\nonumber
+K_5*log(d)*log(h_{\textit{BS}}^{\prime})+K_6*h_{\textit{UE}}^{\prime}\\
+K_7*log(h_{\textit{UE}}^{\prime})+K_{\textit{clutter}}*f(\textit{clutter}).
\end{eqnarray}
Here $K_1=23.8$, $K_2=44.9$, $K_3=10.89$, $K_4=0.19$, $K_5=-10$, $K_6=0$, $K_7=0$, $K_{\textit{clutter}}=1$ are user-defined parameters, $h_{BS}^{\prime}$ and $h_{UE}^{\prime}$ are the effective BS and UE heights, respectively, by taking into account the earth terrain. $L_{\textit{diff}}$ is the diffraction loss calculated by Deygout method and $f(\textit{clutter})$ is the weighted average of the user-specified clutter losses, in the propagation path between BS and UE \cite{Atoll:MCG}.

\subsection{ITU 452 Model}
It is based on the ITU-R P.452-15 recommendation \cite{ITU452} and is valid for 100-500,000 MHz band. It takes into account the LoS/NLoS state, diffraction, tropospheric scatter, surface ducting and elevated layer reflection and refraction. It is given by the following formula:
\begin{eqnarray}
\label{eq:itu1}
\nonumber
L_{\textit{path}}=-5*log\left(10^{-0.2*L_{a}}+10^{-0.2*(L_{b}+(L_{c}-L_{d})*F_j)}\right)\\
+A_{\textit{BS}}+A_{\textit{UE}},
\end{eqnarray}
where,
\begin{eqnarray}
\label{eq:itu2}
\nonumber
F_j=1-0.5*\left[1+tanh\left(2.4*\frac{\theta-0.3}{0.3}\right)\right].
\end{eqnarray}

Here $L_a$ is the basic transmission loss due to troposcatter, $L_b$ is the minimum basic transmission loss with LoS propagation and over-sea sub-path diffraction, $L_c$ is the basic transmission loss associated with diffraction and LoS or ducting/layer-reflection enhancements, $A_{\textit{BS}}$ and $A_{\textit{UE}}$ are additional losses due to BS and UE surroundings, respectively, $F_j$ is the interpolation factor to take into account the path angular distance and $\theta$ is the path angular distance. These parameters are further calculated from equations in ITU-R recommendation P.452-15 \cite{ITU452}.

\subsection{Performance Comparison}
The proposed ML-based propagation model is compared against traditional empirical propagation models, in terms of predictive performance, generalization performance and computational performance.

\subsubsection{Predictive Performance}
\begin{figure}[!h]
        \centering
        \includegraphics[width=1\linewidth]{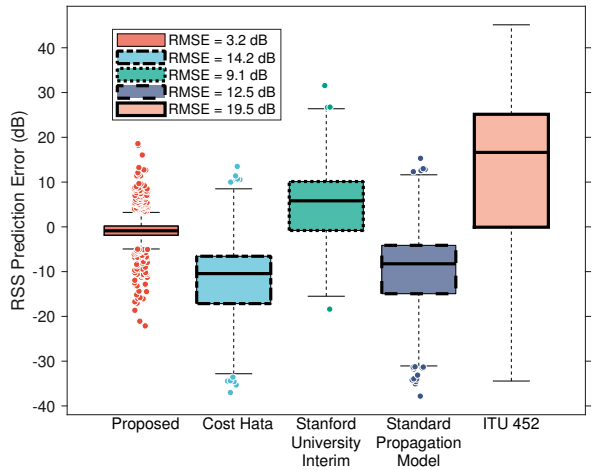}
        \caption{Comparison of Proposed model with various Empirical Radio Propagation Models in terms of RSS prediction error, showing a $65\%$ improvement over the best performing empirical model (3.2 dB RMSE as compared to 9.1 dB for SUI)}
        \label{fig:empirical1}
\end{figure}

In \cref{fig:empirical1}, a box-plot representation is used to compare the performance of our proposed model with the state-of-the-art empirical propagation models, by taking highly precise ray-tracing based RSS estimates as ground truth. The data used here as benchmark is unseen and not used earlier in the training or validation process of our proposed ML-based model. The RSS is calculated from the empirical models using $P_{\textit{UE}}=P_{\textit{BS}}-L_{\textit{path}}$, where $P_{\textit{UE}}$ is the UE's RSS, $P_{\textit{BS}}$ is the BS's transmit power and $L_{\textit{path}}$ is the pathloss calculated using \eqref{eq:cost_hata1}-\eqref{eq:itu1}. We can see that the predicted RSS using our proposed AI-driven model has much less error as compared to other empirical models, showing a $65\%$ improvement over the best performing empirical model (3.2 dB RMSE as compared to 9.1 dB for SUI). 

\subsubsection{Generalization Performance}

The reason for the gain in accuracy of the proposed model lies in its generalizability as compared to other empirical propagation models. Firstly, ML-based model, thanks to its ability to incorporate higher degrees of freedom compared to an empirical, has an intrinsic advantage over empirical model. Empirical models are usually scenario-specific and have different fine-tuned parameter values for different geographic scenarios (e.g., urban, sub-urban, rural etc.) using extensive channel measurements from that scenario. The key cost of this advantage is opaqueness or black box nature of model, which we will address in the next section.

Secondly, through the feature engineering process, the proposed model leverages a novel combination of key features, which are not included in traditional empirical models, and can characterize the physical and geometric structure of the environment traversed by a signal in its propagation path (e.g., indoor distance, Manhattan distance, number of building penetrations in each clutter type etc.), and are sensitive to the change in network parameters (e.g., horizontal angular separation, vertical angular separation etc.). These additional features (or degrees of freedom) enable the ML-model to be trained on combined data from different geographic scenarios and hence provide more scalability and generalizability.

\begin{figure}[!h]
        \centering
        \includegraphics[width=1\linewidth]{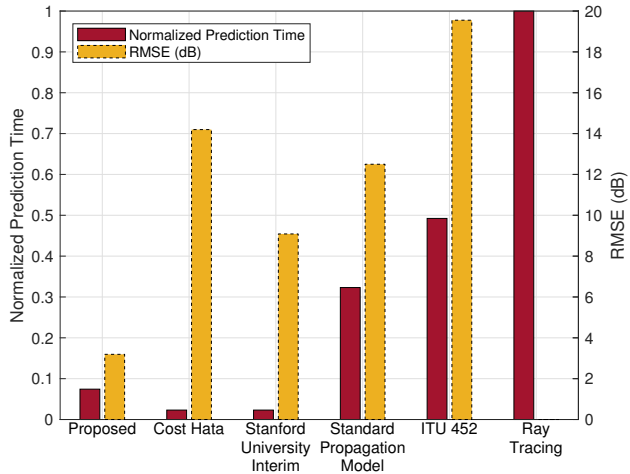}
        \caption{Comparison of Proposed model with various Empirical Radio Propagation Models and Ray Tracing w.r.t. Prediction Time and RSS Prediction Error. The proposed ML-based model is atleast 65\% more accurate than the investigated empirical models, and 13x faster than ray-tracing}
        \label{fig:empirical2}
\end{figure}

\subsubsection{Computational Performance}
While the proposed model yields better accuracy than empirical models, our analysis (shown in \cref{fig:empirical2}) shows that its computational complexity and therefore implementation cost is much lower than the highly sophisticated ray-tracing based tools that are being widely used in commercial cell planning tools, because it only uses the key features as input to the trained ML-based model to predict the RSS, as compared to ray tracing, which approximates the interactions of all rays with the neighboring environment to estimate the pathloss, hence computationally inefficient. As a result, it's much faster than ray-tracing, and thereby addresses a much-complained problem in ray-tracing based tools, by industry professionals. The preliminary implementation of the proposed framework has demonstrated a 13x decrease in prediction time as compared to ray-tracing approach, and can be further optimized to make it more efficient (for instance, by using parallel computing).

\section{Secondary Analysis for Interpretability and Sensitivity}\label{section4}
One of the key caveats of applying machine learning is lack of interpretability of the resultant models. This challenge often undermines the uptake rate of ML based models, particularly in cellular networks where stakes are high. Therefore, knowing why a model is predicting what it is predicting can be a very useful auxiliary information on top of accuracy, prediction and training agility and robustness to sparsity of training data. Model interpretation is  also a vital debugging tool, as it can help you learn about the problems (e.g., biasness) in the model and for ensuring that small changes in the input do not lead to large changes in the prediction. Therefore, in this section, we try to make our proposed black-box machine learning model more trustable, interpretable and robust \cite{doshi2017towards}.

\subsection{Sensitivity Analysis}
Sensitivity Analysis is a useful technique for investigating the model's behavior for specific scenarios of interest and for providing a global insight into the model's behavior. This is done by quantifying the contribution of each input feature, in the variability of the model output. These values are called \textit{sensitivity indices}.

\subsubsection{Sobol Indices}
The most popular method of finding these sensitivity indices is \textit{Sobol Method} \cite{sobol1993sensitivity}, which is based on the variance of model output. However, they are very difficult to interpret if there is a statistical dependence between features. For Instance, in the case of independent features, there exists a unique \textit{Sobol Index} for a feature, representing the variance in model output solely by that feature, also called \textit{First Order Sobol Index}. But if the features have dependency between them, then the first order indices fail to capture the contribution of each feature, and \textit{Second Order Sobol Indices} are used to express the contributions of the interactions between each pair of features, and so on for higher orders.

\subsubsection{LOCO Variable Importance}
Leave-one-covariate-out (LOCO), or even Leave-one-feature-out (LOFO) \cite{lei2018distribution} is another method for finding variable importance (or sensitivity) in the model output. It scores each row in the training data for each feature (or covariate). In each scoring run, one feature is missed and its impact on the output prediction is measured. The feature with the most impact on the predicted outcome is taken to be the most important. However, its performance can quickly deteriorate if there are complex non-linear dependencies in a model, in which case Shapley values will be a better technique.

\subsubsection{Shapley Values}
The lack of accurate model interpretation using the above methods, when there are complex non-linear interdependencies between features, can be overcome by using \textit{Shapley values} \cite{shapley1953value}, which is a Nobel-laureate concept in cooperative game theory and economics, to determine the contribution of each player in a collaborative game to its success, but can be used to calculate feature importance in a model and thus achieve a good degree of interpretability, even for non-parametric models. \textit{In the case of dependence between a group of input features, the effect of interaction between features is equally allocated to each feature within the group}.

\subsection{Model Interpretation with SHAP}
A recently introduced method called \textit{SHAP (SHapley Additive exPlanations)} \cite{lundberg2017unified}, based on Shapley values, measures how much each feature contributes, either positively or negatively, to the model output. An advantage of using SHAP is that each sample in the data has its own set of SHAP values, unlike traditional methods, which only tells the importance of a feature across the whole dataset. This is particularly useful, as we can observe the effect on model output, for the whole range of each input feature. In our further analysis, we have used the TreeSHAP algorithm \cite{TreeSHAP}, which is an efficient approach of calculating shapley values of \ac{ML} models belonging to decision tree family (e.g., LightGBM, XGBoost etc.).\\

\subsubsection{Feature Importance using SHAP Summary Plots}
\begin{figure*}[!t]
        \centering     
        \subfigure[SHAP Value Distribution]{\includegraphics[width=.49\linewidth]{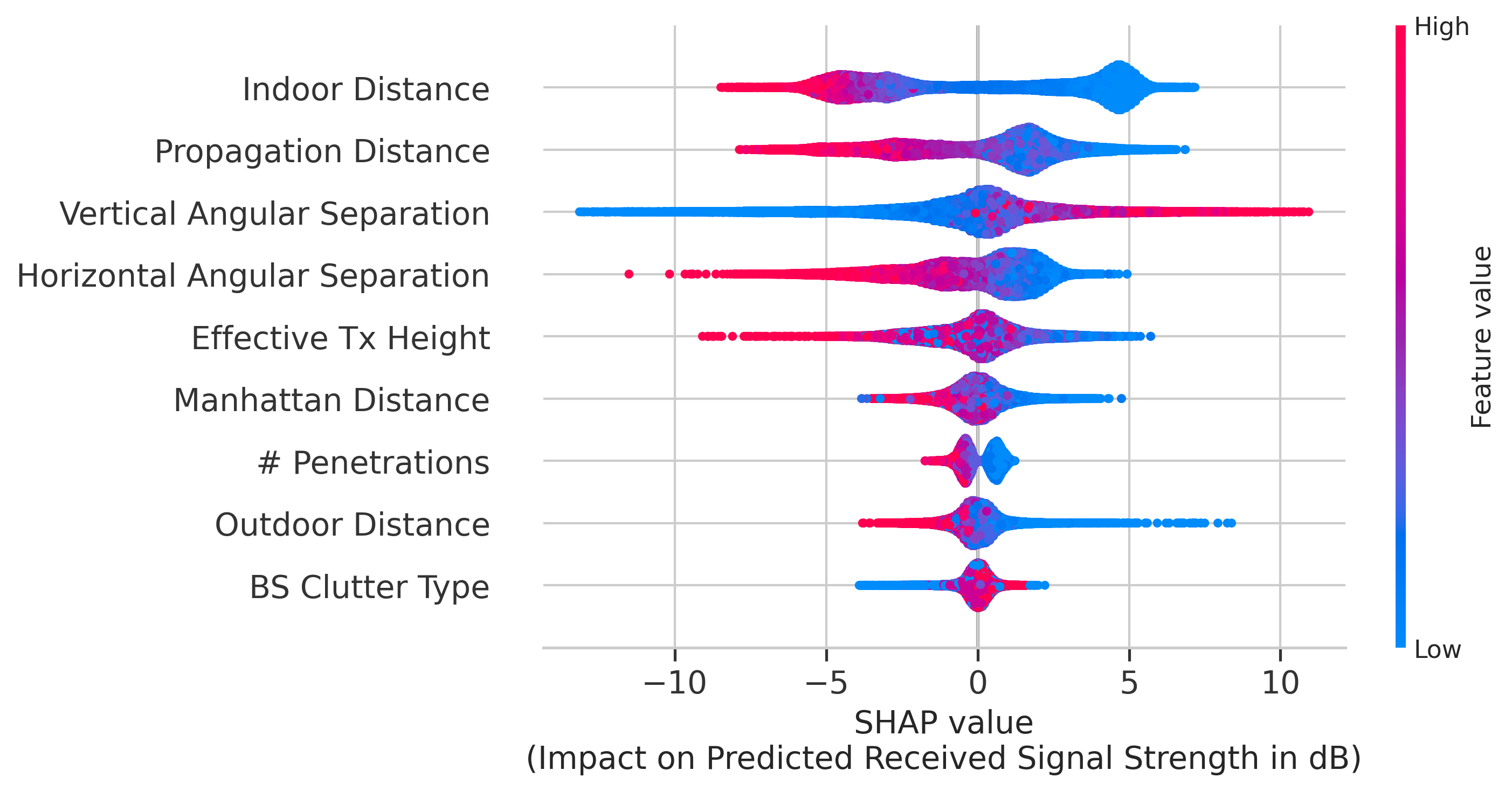}\label{fig:shap1a}}
        \subfigure[Mean SHAP Value]{\includegraphics[width=.49\linewidth]{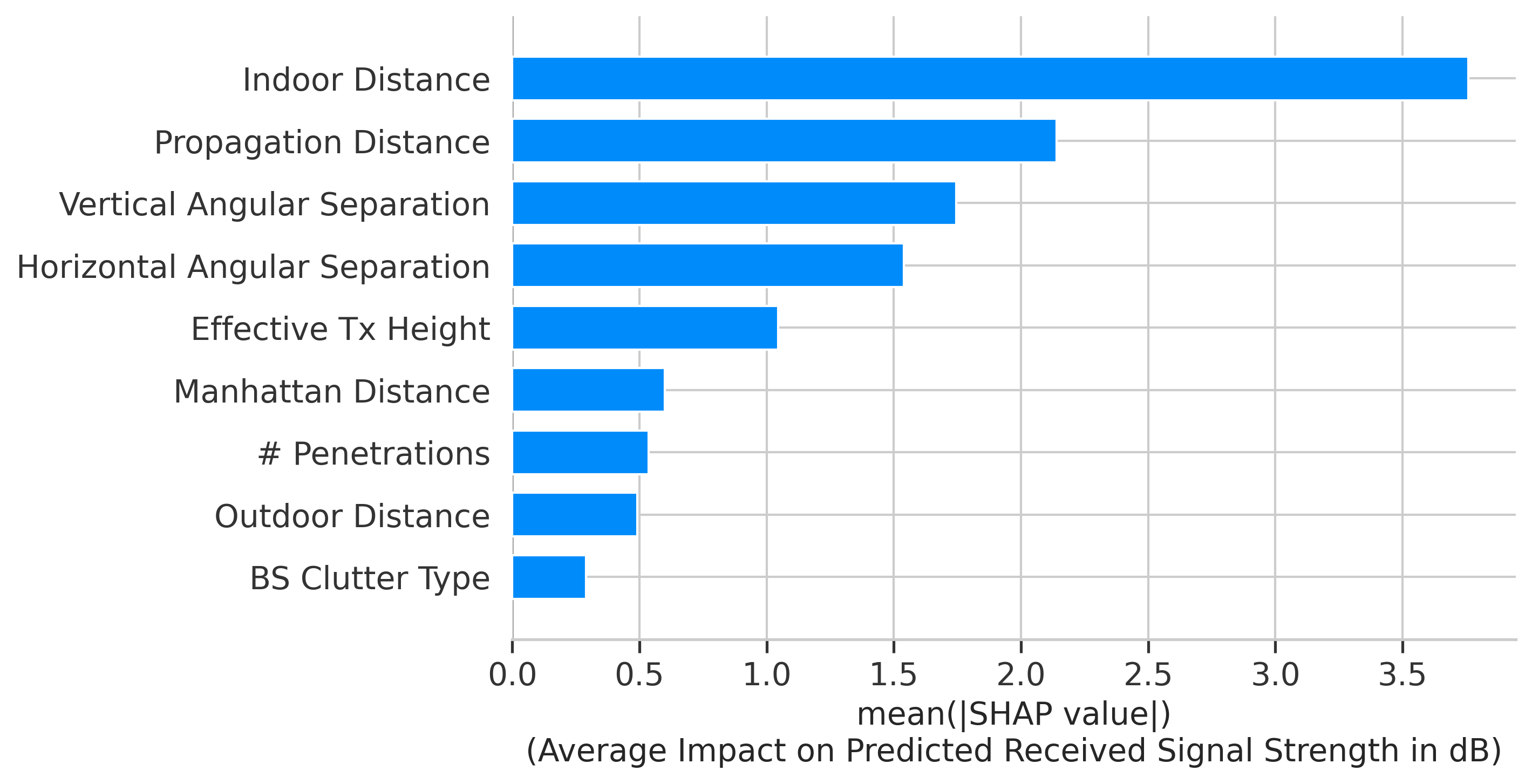}\label{fig:shap1b}}
        \caption{({\bf a}) SHAP Summary plots showing Top 9 Input Features ({\bf a}) SHAP Value (Impact on RSS) variance w.r.t respective feature values ({\bf b}) Mean SHAP Values (Average Impact on RSS)}
        \label{fig:shap1}
\end{figure*}

\begin{figure*}[!ht]
        \centering     
        \subfigure[]{\includegraphics[width=.32\linewidth]{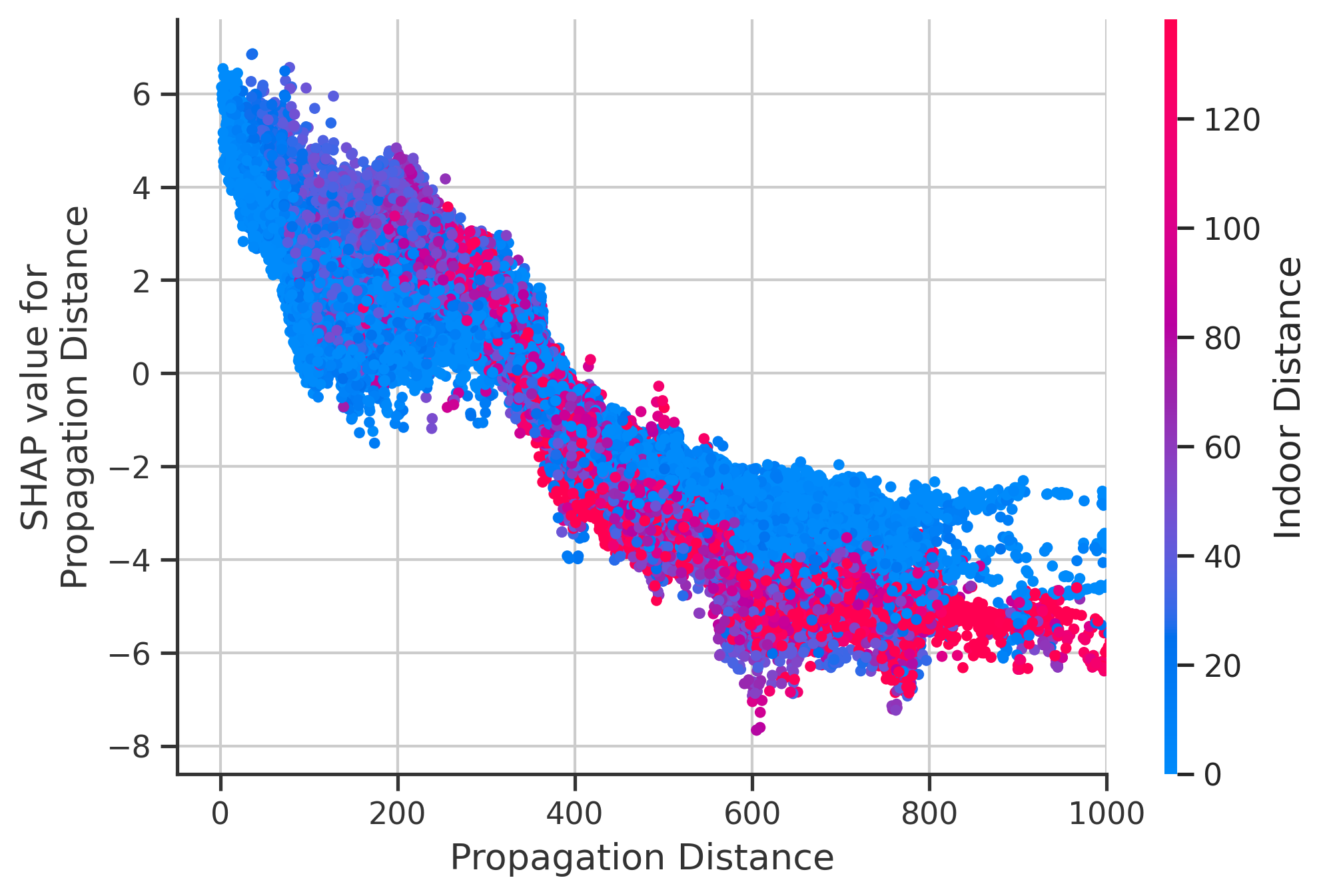}\label{fig:shap2a}}
        \subfigure[]{\includegraphics[width=.32\linewidth]{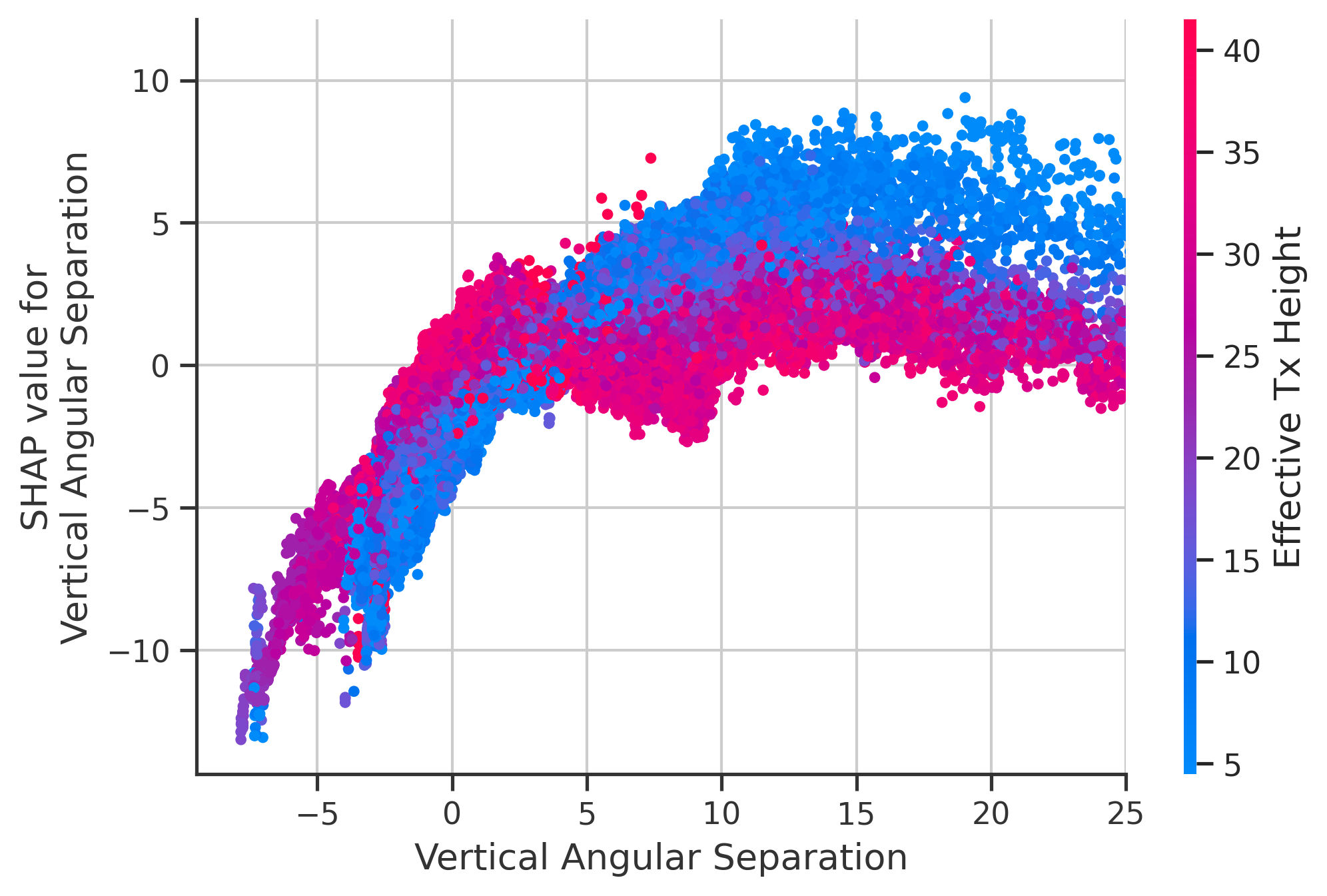}\label{fig:shap2b}}
        \subfigure[]{\includegraphics[width=.32\linewidth]{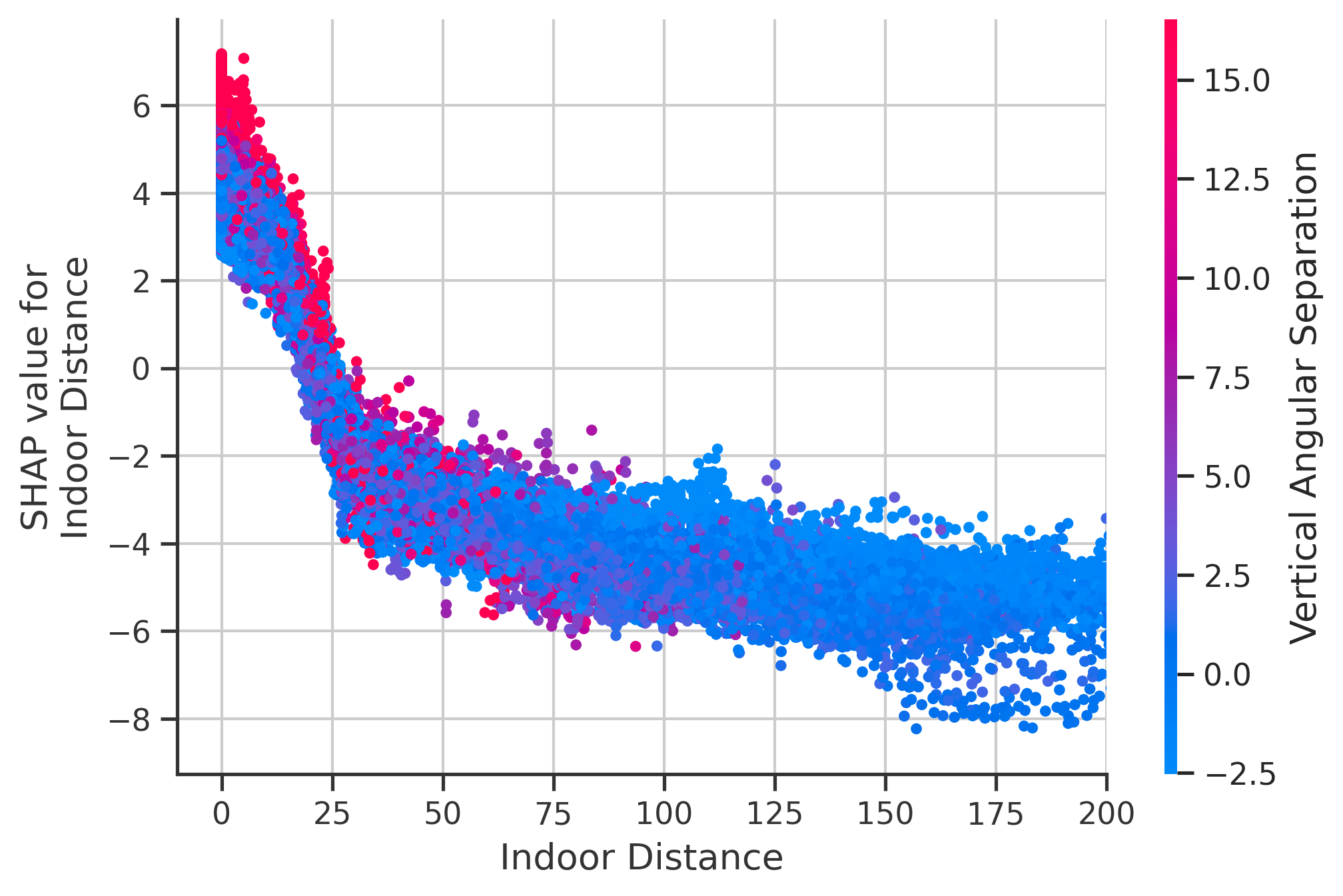}\label{fig:shap2c}}
        \subfigure[]{\includegraphics[width=.32\linewidth]{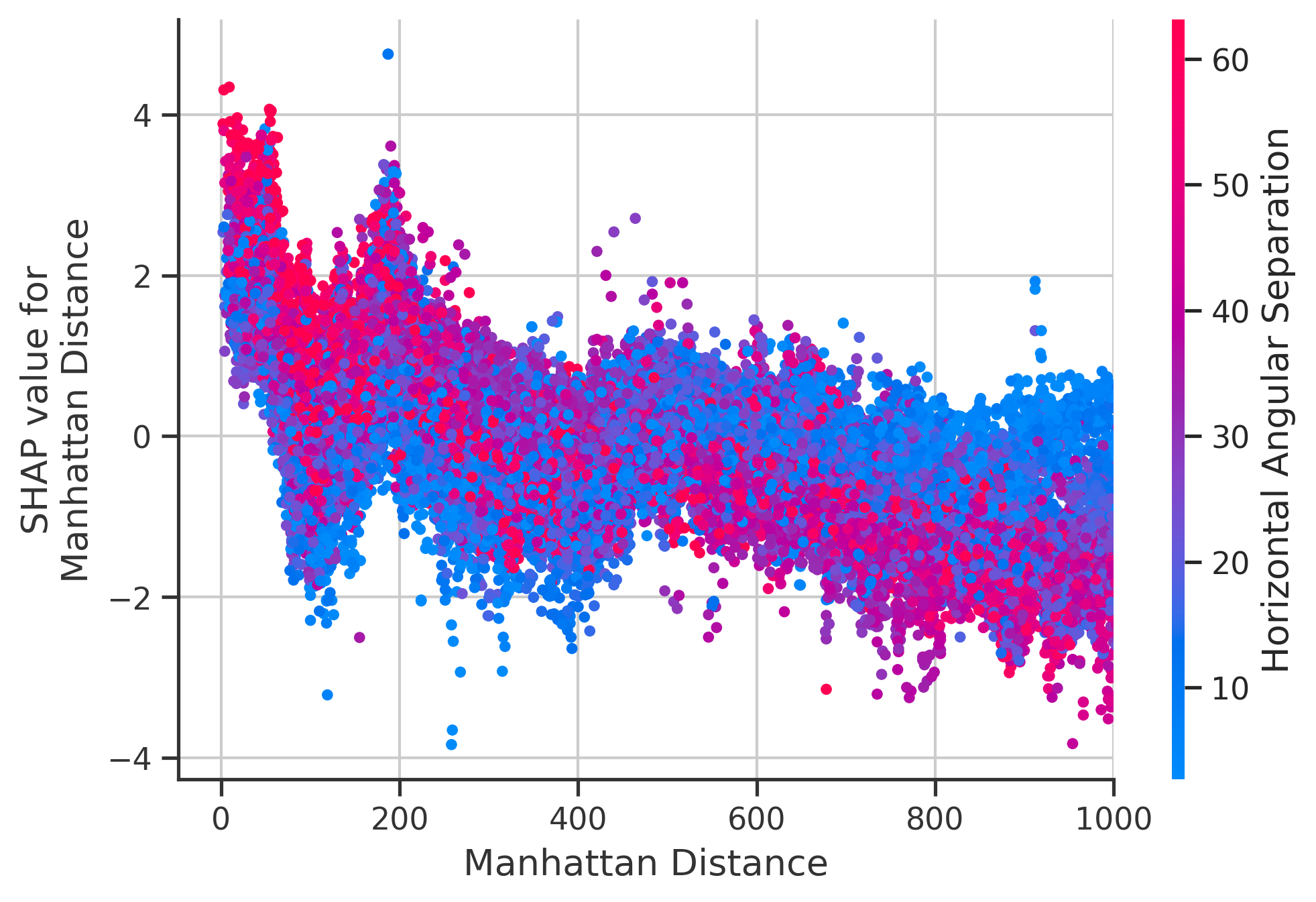}\label{fig:shap2d}}
        \subfigure[]{\includegraphics[width=.32\linewidth]{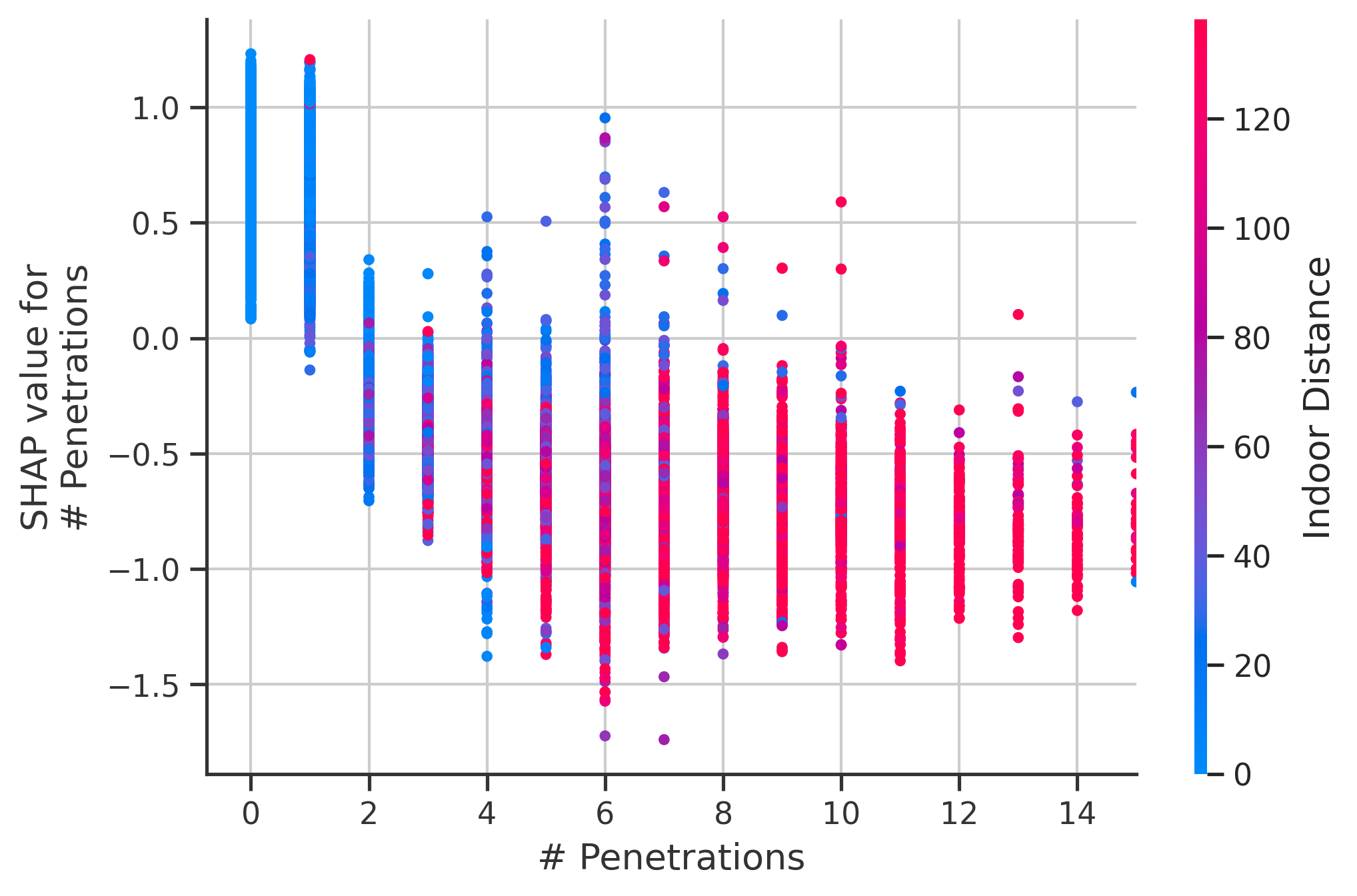}\label{fig:shap2e}}
        \subfigure[]{\includegraphics[width=.32\linewidth]{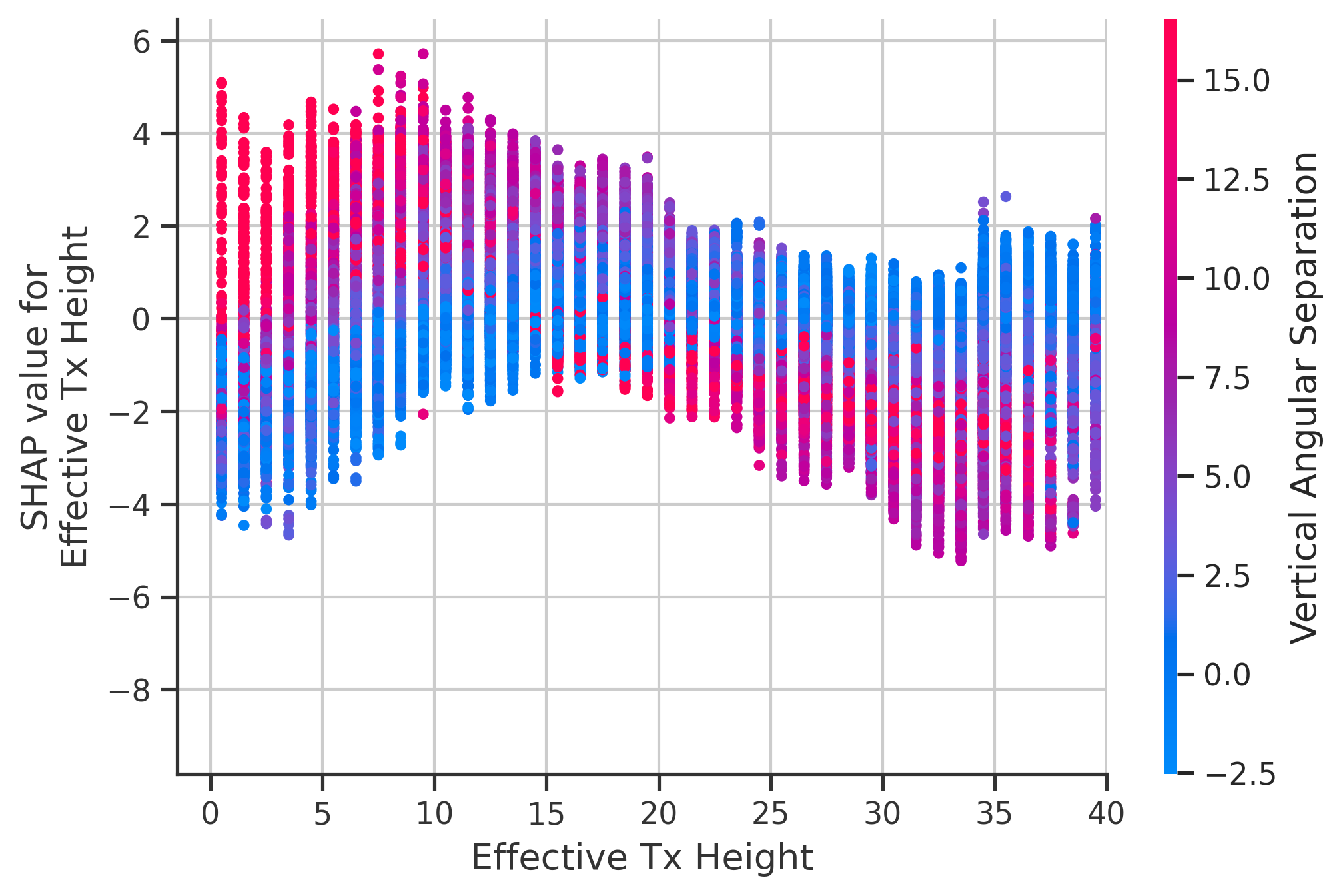}\label{fig:shap2f}}
        \caption{Impact of various features and their inter-dependence in predicting Received Signal Strength. Useful for model interpretability and finding important feature regions (or values) for intelligent data collection and network automation}
        \label{fig:shap2}

\end{figure*}

In \cref{fig:shap1a}, SHAP values of some features are plotted for all measurement instances, which show the distribution of impacts on the predicted RSS value, for each input feature. Here the \textit{points are colored by the respective feature's value and piled up vertically to show density}. For each measurement sample, the sum of SHAP values (for every feature) equals the variance in the predicted output from its mean value across all samples.  For instance, from domain knowledge we know that the RSS of a user will be highest if its \textit{horizontal angular separation} from the BS antenna is close to zero and vice versa, due to the impact of antenna beamwidth on its attenuation. Similarly, the RSS of a user will be high if it's close to the BS, therefore, the impact of \textit{propagation distance} is highest at its extreme values, same is the case for \textit{indoor distance} and \textit{outdoor distance}. On the other hand, \textit{Vertical Angular Separation} is generally inversely proportional to the distance between BS and UE, therefore its impact is highest when it has a high value and vice versa. Also the impact of \textit{Effective BS Height} is high, if the net vertical distance between the UE and BS is high, and vice versa. Similar trends can be seen for other features as well.

In \cref{fig:shap1b}, the mean SHAP value for each input feature is plotted. These results show the average impact of each feature on the model output (i.e., predicted RSS value). We can see that, contrary to the common understanding where distance is considered and used in literature as the key determining factor for pathloss (or RSS), the \textit{Indoor Distance} has the highest feature importance (or impact) in the RSS model. On the other hand, \textit{BS Clutter Type} in the propagation path has the lowest impact.

\subsubsection{Feature Inter-Dependency using SHAP Dependency Plots}
The interplay of combinations of features can be uncovered using SHAP dependence plots. By plotting the SHAP value for many samples in the dataset (See \cref{fig:shap2}), we can see that the SHAP value (attributed importance) of a feature changes as its value varies. However, its interaction with other features in the model is captured by its vertical dispersion. Unlike \textit{standard partial dependence plots}, that only plot a line, here each dot (sample) is colored with the value of an interacting feature.

In \cref{fig:shap2a}, we can see that the impact of \textit{propagation distance} decreases as its value increases. Whereas, as mentioned before, \textit{indoor distance} has an interaction with the propagation distance that affects its relative importance. \cref{fig:shap2b} shows the effect of \textit{effective BS height} on the attributed feature importance of \textit{vertical angular separation}, where high value of effective BS height decreases the importance of vertical angular separation when its value is greater than zero, and vice versa. Similarly, in \cref{fig:shap2c}, we can see the increase in feature importance of \textit{indoor distance} at points where vertical angular separation is high. \cref{fig:shap2d} shows that feature importance of \textit{Manhattan distance} and its interplay with horizontal angular separation. In \cref{fig:shap2e}, as we know that the increase in \textit{number of building penetrations} in the propagation path between a BS and UE, increase its indoor distance as well in most cases, results in the decrease of its feature importance (\cref{fig:shap2e}). Lastly, \cref{fig:shap2f} shows the feature importance of effective BS height and its interaction with vertical angular separation.

\subsection{Insights from Interpretability/Sensitivity Analysis}
To interpret the model predictions and gain insights into the black-box model by turning it into rather grey-box model, SHAP algorithm is used. The SHAP Summary plot (or the feature importance plot) in \cref{fig:shap1} shows the mean importance of each feature in the variability of the model output. This plot is particularly useful for a system-level control as it shows that what control knob (or network parameter) needs to be played the most for tuning network configuration to get optimal performance.
SHAP Dependency plots (shown in \cref{fig:shap2}), on the other hand, shows the behavior of feature importance (or SHAP value) with respect to the value of its corresponding feature and its interaction with the most dependent feature. This plot is useful for observing the range of values for a pair of features that has the highest impact on the model output. For Example, \cref{fig:shap2c} shows that the \textit{indoor distance} and \textit{vertical angular separation} have the highest impact on the model output when 0 < $d_{\textit{indoor}}$ < 20 m and $\phi_{\textit{ver}} > 10^{\circ}$.
\subsection{Utility of Insights Gained from the Proposed Model}
The information yield by the SHAP analysis, that has transformed the originally black-box model into a \emph{grey-box model}, can be exploited in real networks for several use cases. Below we identify three key use cases: 

\begin{enumerate}
\item \uline{Addressing the Sparsity Challenge:} A key challenge in applying ML to wireless networks is sparsity of training data i.e., gathering data for complete parameters ranges is often very difficult, if not impossible. For example, its not viable to gather RSS measurements against all antenna tilt range (0-90) in a live network. Furthermore, usually the process of gathering and enriching training data is costly. The proposed framework builds a grey-box model instead of a black box model, thanks to the insights provided by the SHAP analysis, can be leveraged to address the aforementioned challenges of data sparsity. The knowledge that what parameter ranges are more crucial to the model can be exploited for selective collection and enrichment of training data. This can provide a lower cost/benefit ratio as compared to a uniform or random collection or enrichment of training data. For example, based on observation from \cref{fig:shap2c}, instead of uniform or random measurement campaigns, more resources should be dedicated to data collection for Antenna Tilt and UE's RSS data pairs corresponding to $\phi_{\textit{ver}} > 10^{\circ}$ (vertical angular separation) and Antenna Tilt/Azimuth and UE's RSS data pairs corresponding to 0 < $d_{\textit{indoor}}$ < 20 (indoor distance in the propagation path).

\vspace{5pt}
\item \uline{Intelligent Optimization:} Current design and post deployment optimization paradigm of cellular networks rely mostly on the domain knowledge.  However, given the large number of design and optimization parameters per site\textemdash already roughly 1500/site in LTE \textemdash\ and growing complexity trend towards 5G and beyond, achieving optimal design and operation in emerging cellular networks by solely relying on domain knowledge is going to become inviable approach. The insights gained from the semi transparency (vis-a-vis greyness) of the presented model achieved through the proposed framework can be very helpful towards more effective and resource efficient design and post deployment optimization of the network. For example, while searching for optimal design and configuration parameters, the parameters and regions of the search space with parameter ranges identified by the proposed framework to be more influential on the KPIs, can be explored more exhaustively, compared to other parameters and parts of the parameter range. This  approach is expected to improve the design and optimization processes compared to uniform (brute force based) or pseudo random or heuristic search algorithms (e.g.,  genetic algorithms, simulated annealing) based design and optimization.

\vspace{5pt}
\item \uline{Lighter ML model for low-latency use-cases:} The insights gained from the SHAP analysis can also be used to select the most important features for building our proposed ML model. Therefore, a lighter version of the model can be built using the selected key features to further reduce the computational complexity of the model. The results from this analysis (in \cref{tab:lighter_ml}) show that by using the top 5 features (from \cref{fig:shap1b}), i.e., indoor distance, propagation distance, vertical angular separation, horizontal angular separation and effective BS height, the training and prediction time of the resulting \ac{RSS} prediction model can be significantly reduced (by around 70\%) at the cost of negligible loss in accuracy (by around 3\%) to enable low latency use-cases for the proposed SHAP-enabled lighter model. By allowing real-time predictions, such lightweight model can be used for real time AI-powered closed loop optimization of the network, thus acting as a key enabler for the \acf{URLLC} in 5G networks.

\begin{table}[!h]
        \begin{center}
                \caption{Performance Evaluation of the SHAP-enabled Lighter Model using 5-fold Repeated Cross Validation}
                \label{tab:lighter_ml}
                \renewcommand{\arraystretch}{1.3}
                \begin{tabular}{|c|c|c|c|}
                        \hline
                        \thead{Performance \\Metric} & \thead{Baseline ML \\Model using \\all features} & \thead{Lighter ML\\ Model using \\Top 5 features} & \thead{Net\\Gain}\\
                        \hline
                        RMSE & $3.542 \pm 0.036$ & $3.661 \pm 0.038$ & $-3.35\%$\\
                        \hline
                        $R^2$ & $0.854 \pm 0.003$ & $0.844 \pm 0.003$ & $-1.17\%$\\
                        \hline
                        Training Time & $1.896 \pm 0.090$ & $0.681 \pm 0.049$ & $\mathbf{+64.07\%}$\\
                        \hline
                        Prediction Time & $0.077 \pm 0.005$ & $0.023 \pm 0.002$ & $\mathbf{+70.13\%}$\\
                        \hline
                \end{tabular}
        \end{center}
\end{table}

\end{enumerate}

\section{Conclusion and Future Works}\label{section5}
In this paper, we propose a framework for a robust and scalable AI-driven 3D propagation model for cellular networks. To enable this framework, we have identified a novel set of key predictors, that can characterize the complex physical and geometric structure of the propagation environment. Performance comparison of several state-of-the-art machine learning algorithms including Linear Regression, K-Nearest Neighbors, Decision Tree, Random Forest, Extremely Randomized Trees, Adaptive Boosting (AdaBoost), Gradient Boosting Decision Trees (GBDT), Extreme Gradient Boosting (XGBoost), Light Gradient Boosting (LightGBM), Categorical Boosting (CatBoost) and Deep Neural network (DNN) is done to highlight their strengths and weaknesses in modeling the propagation through complex real environment using the proposed key predictors as input features. The results show that \textit{LightGBM} outperforms other ML tools, including DNN, in terms of computational complexity and robustness to extremely sparse training data (just 2\%), as often is the case in real networks. On the other hand, compared to other tested ML tools, DNN's performance deteriorates the most when the training data becomes sparse. The proposed ML-Based model is compared against state-of-the-art empirical models including COST-Hata, Stanford University Interim (SUI), Standard Propagation Model (SPM) and ITU 452 Model. Proposed ML-based model yields 65\% higher accuracy in RSS estimation as compared to empirical propagation models, when highly sophisticated ray-tracing based data for the city of Belgium from a commercial planning tool is used as ground truth. On the other hand, proposed model offers 13x reduction in prediction time as compared to ray-tracing based commercial planning tool. The black box nature of the proposed model is transformed into relatively more interpretable grey-box model using SHAP method. The insights presented through interpretability analysis offer new research directions such as intelligent data gathering for addressing the challenge of training data sparsity, finding the optimal combination of network configuration parameters and building lighter ML models for low-latency use-cases.

The presented system level pathloss prediction model combined with the interpretability analysis thus manifests a framework that can act as a corner stone for the Artificial Intelligence driven Self Organizing Networks (AISON), as opposed to current SON which lacks interpretable models for quantifying network performance as function of the plethora of network configuration parameters. Furthermore, in addition to system level pathloss model, the proposed framework can be extended to an AI-driven link level channel model for channel estimation, physical layer design etc. Such extension will require incorporating many other channel parameters such as delay spreads and angular spreads etc., that can be neglected in system level path lass model due to the much larger temporal and spatial scale that can suffice for system level planning and optimization, but cannot give meaningful insights for link level design. Other possible extensions are developing both link level and system level channel models for higher frequency bands. This is needed especially for next-generation 3D heterogeneous cellular networks in mmWave/Terahertz bands with flying \acf{UAV}-based \acp{BS} \cite{zhang2019survey,qureshi2019tradeoffs} and \acf{RIS} \cite{basar2019wireless}, since propagation conditions will be significantly different than sub-6 GHz bands and conventional network planning using empirical model will cease to be a viable option.

\bibliographystyle{IEEEtran}
\bibliography{bibliography}

\begin{IEEEbiography}[{\includegraphics[trim={4cm 0 4cm 0}, clip, width=1in,height=1.25in, keepaspectratio]{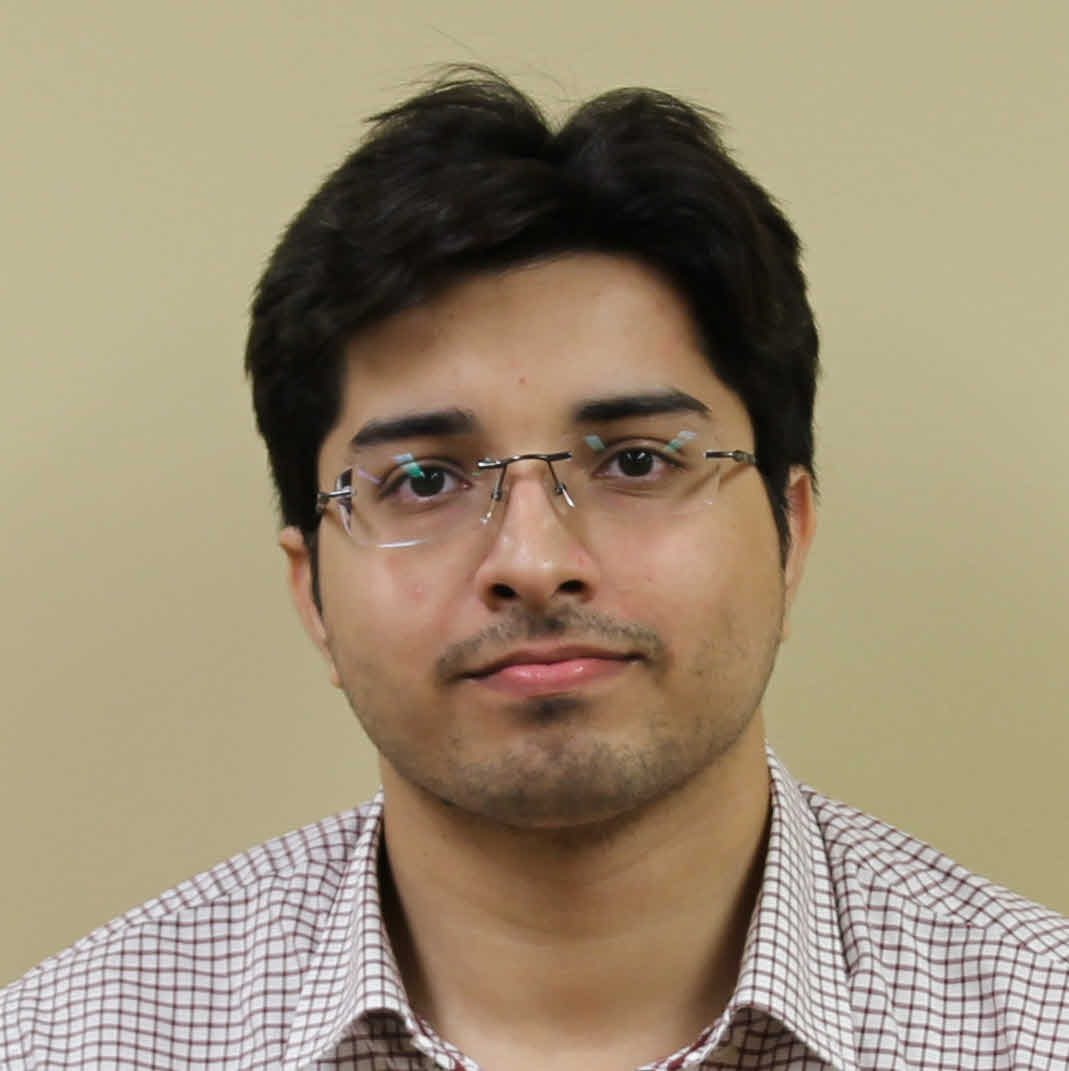}}]{Usama Masood} is currently working toward his Ph.D. degree in electrical and computer engineering at the Artificial Intelligence for Networks Center, University of Oklahoma (OU), Tulsa, OK. His research focus at AI4Networks Center is on designing AI-enabled system-level solutions for Zero Touch Automation in emerging networks. During his PhD, he has contributed to several National Science Foundation (NSF) research and development projects on 5G Cellular Networks including the design and development of 5G Testbed at OU-Tulsa campus for enabling experimental research on different 5G use-cases and designing advanced cell outage management framework for self-healing networks, a multi-national project on AI-based Self Organizing 5G and beyond networks and a industry collaboration project with Ericsson Research, CA for solving the training data sparsity challenge to enable AI-based zero touch automation in emerging networks.
\end{IEEEbiography}

\begin{IEEEbiography}[{\includegraphics[trim={0 0 3cm 0}, width=1in,height=1.25in,clip,keepaspectratio]{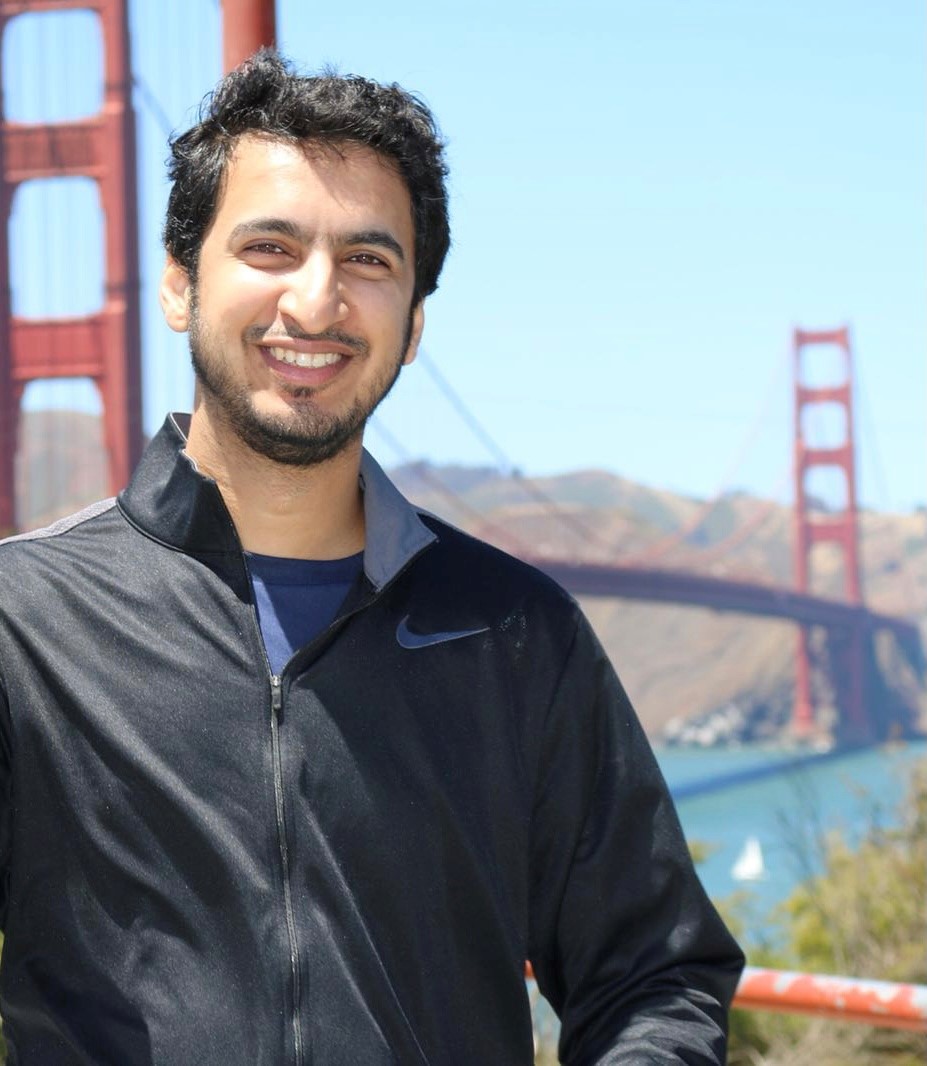}}] {Dr. Hasan Farooq} received his B.Sc. degree in Electrical Engineering from the University of Engineering and Technology, Lahore, Pakistan, in 2009, M.Sc. by Research degree in Information Technology from Universiti Teknologi PETRONAS, Malaysia in 2014 and his Ph.D. degree in Electrical and Computer Engineering from the University of Oklahoma, USA in 2018. Hasan has developed several methods to harness unprecedented intelligence from the wireless communication network data that can be utilized for proactive dynamic network orchestration and optimization. His dissertation was awarded Gallogly College of Engineering Dissertation Excellence Award by University of Oklahoma in 2018. He has contributed to several NSF funded research projects in the domain of self-organizing networks for 5G and beyond and has authored over 40 high impact publications. He is currently working as AI Researcher at Ericsson Research Silicon Valley.
\end{IEEEbiography}

\begin{IEEEbiography}[{\includegraphics[width=1in,height=1.25in,clip,keepaspectratio]{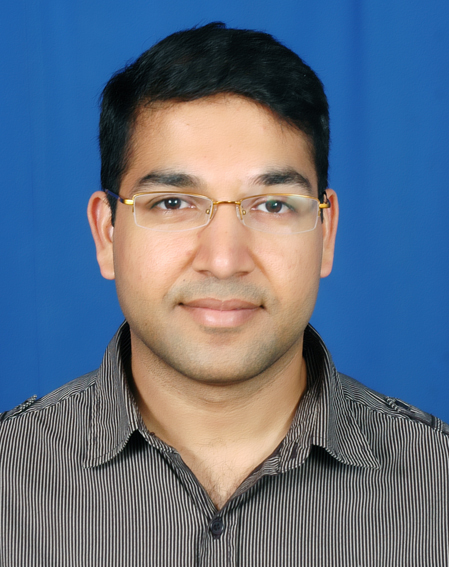}}] {Ali Imran} (Senior Member, IEEE) is a Presidential Associate Professor and Willams Professor in ECE and founding director of the Artificial Intelligence (AI) for Networks (AI4Networks) Research Center and TurboRAN Testbed for 5G and Beyond at the University of Oklahoma. His research interests include AI and its applications in wireless networks and healthcare. His work on these topics has resulted in several patents and over 100 peer-reviewed articles including some of the highly influential papers in the domain of wireless network automation. On these topics he has led numerous multinational projects, given invited talks/keynotes and tutorials at international forums and advised major public and private stakeholders and co-founded multiple start-ups. He holds a B.Sc. degree in electrical engineering from the University of Engineering and Technology Lahore, Pakistan, in 2005, and the M.Sc. degree (Hons.) in mobile and satellite communications and the Ph.D. degree from the University of Surrey, Guildford, U.K., in 2007 and 2011, respectively. He is an Associate Fellow of the Higher Education Academy, U.K. He is also a member of the Advisory Board to the Special Technical Community on Big Data, the IEEE Computer Society.
\end{IEEEbiography}

\begin{IEEEbiography}[{\includegraphics[trim={1cm 0 0 0}, width=1in,height=1.25in,clip,keepaspectratio]{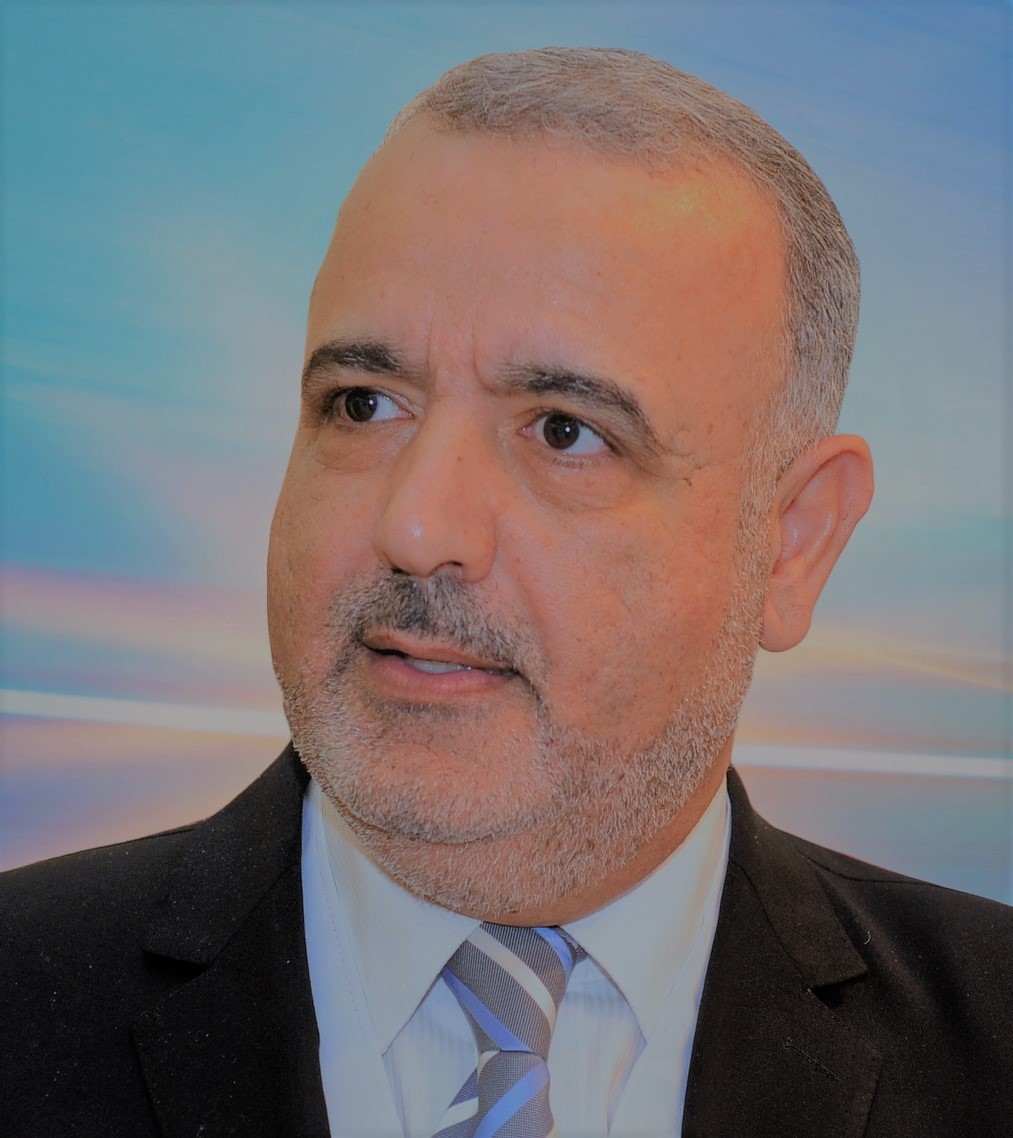}}] {Dr. Adnan Abu-Dayya} has been serving as the Founding Executive Director of the Qatar Mobility Innovations Center (QMIC) since 2009. It is one of the first technology innovation institutions in the Middle East focused on translating R\&D and technology innovations into scalable digital platforms and solutions in the field of Intelligent Mobility and Smart Cities. From 2007 to 2008, Adnan served as the Chairman of the Electrical Engineering Department at Qatar University. Before moving to Qatar in 2007, Adnan worked for 10 years at AT\&T Wireless in Seattle, USA where he served in a number of senior management positions covering product innovations, emerging technologies, systems engineering, product realization, and intellectual property management. Before that, Adnan worked as a Senior Manager at Nortel Networks in Ottawa, Canada in the advanced technology group, and as a Senior Consultant at the Communications Research Centre in Ottawa, Canada. 
Dr. Adnan serves as the Chairman of the Advisory Board of the Electrical \& Computer Engineering Department of Texas A\&M University at Qatar,  he is a member of the Steering Committee of the Smart Grid Research Center at Texas A\&M University at Qatar, and is a member of the advisory board of the Qatar Car Museum. Adnan received his PhD in Wireless Communications (Electrical Engineering) from Queens University, Kingston, Canada in 1992. He has 10 issued patents, and more than 100 referred publications.

\end{IEEEbiography}

\end{document}